**Title**

# Calibration of the NDHA model to describe N2O dynamics from respirometric assays.

**Author list**

Carlos Domingo-Félez[a], María Calderó-Pascual[a], Gürkan Sin[b], Benedek G. Plósz[a,1], Barth F. Smets[a]*

[a]Department of Environmental Engineering, Technical University of Denmark, Miljøvej 115, 2800 Kgs. Lyngby, Denmark

[b]Department of Chemical and Biochemical Engineering, Technical University of Denmark, Søltofts Plads 229, 2800 Kgs. Lyngby, Denmark

[1]Present address: Department of Chemical Engineering, University of Bath, Bath, England.

* Corresponding author:

Barth F. Smets, Phone: +45 4525 1600, Fax: +45 4593 2850, E-mail: bfsm@env.dtu.dk

**Abstract**

The NDHA model comprehensively describes nitrous oxide ($N_2O$) producing pathways by both autotrophic ammonium oxidizing and heterotrophic bacteria. The model was calibrated via a set of targeted extant respirometric assays using enriched nitrifying biomass from a lab-scale reactor. Biomass response to ammonium, hydroxylamine, nitrite and $N_2O$ additions under aerobic and anaerobic conditions were tracked with continuous measurement of dissolved oxygen (DO) and $N_2O$.

The sequential addition of substrate pulses allowed the isolation of oxygen-consuming processes. The parameters to be estimated were determined by the information content of the datasets using identifiability analysis. Dynamic DO profiles were used to calibrate five parameters corresponding to endogenous, nitrite oxidation and ammonium oxidation processes. The subsequent $N_2O$ calibration was not significantly affected by the uncertainty propagated from the DO calibration because of the high accuracy of the estimates. Five parameters describing the individual contribution of three biological $N_2O$ pathways were estimated accurately (variance/mean < 10% for all estimated parameters).

The NDHA model response was evaluated with statistical metrics (F-test, autocorrelation function). The 95% confidence intervals of DO and $N_2O$ predictions based on the uncertainty obtained during calibration are studied for the first time. The measured data fall within the 95% confidence interval of the predictions, indicating a good model description. Overall, accurate parameter estimation and identifiability analysis of ammonium removal significantly decreases the uncertainty propagated to $N_2O$ production, which is expected to benefit $N_2O$ model discrimination studies and reliable full scale applications.

**Keywords**: Nitrous oxide, Respirometry, Accuracy, Modelling, Parameter Estimation

**Highlights**

- An experimental design to calibrate $N_2O$ models via extant respirometry is proposed.
- Parameters associated to $N_2O$ production were identified accurately (CV < 10%).
- The NDHA model described $N_2O$ production at varying DO and $HNO_2$ concentrations.
- The uncertainty of the $N_2O$ emission factor in a case study was ~ 10% of its value.

## 1. Introduction

Nitrous oxide ($N_2O$) is a greenhouse gas emitted during biological nitrogen removal (BNR). The carbon footprint of wastewater treatment plants (WWTPs) is highly sensitive to $N_2O$ emissions (Gustavsson and Tumlin, 2013), thus reducing $N_2O$ emissions is beneficial for the energy balance of WWTPs.

During BNR operations biological and abiotic pathways are responsible of $N_2O$ emissions (Schreiber et al., 2012). Ammonia-oxidizing bacteria (AOB) produce $N_2O$ during incomplete ammonium ($NH_4^+$) oxidation to nitrite ($NO_2^-$) (nitrifier nitrification, NN). Under low dissolved oxygen (DO) AOBs use $NO_2^-$ as the terminal electron acceptor and also release $N_2O$ (nitrifier denitrification, ND). Heterotrophic denitrification is a 4-step process where $N_2O$ is an obligate intermediate. Under low carbon-to-nitrogen ratios or in the presence of DO heterotrophic denitrification is not complete and $N_2O$ can be released (HD) (Richardson et al., 2009). Hydroxylamine ($NH_2OH$) and free nitrous acid ($HNO_2$) are intermediates of $NH_4^+$ oxidation which can produce $N_2O$ abiotically.

Process models are useful tools that translate our understanding of $N_2O$ production into mathematical equations. $N_2O$ model structures vary depending on the number of pathways, nitrogenous variables or parameters considered (Ding et al., 2016; Ni et al., 2014; Pocquet et al., 2016). The description of the autotrophic contribution transitioned form single- (NN or ND) to two-pathway (NN and ND) models to capture the $N_2O$ dynamics observed during N-removal (Pocquet et al., 2016; Schreiber et al., 2009). Recently, a consilient $N_2O$ model was proposed (NDHA) that predict three biological pathways and abiotic processes (Domingo-Félez and Smets, 2016). Potentially, the NDHA model describes $N_2O$ production under a wide range of operational conditions.

$N_2O$ models are extensions of existing structures describing nitrogen removal and thus, calibration of $N_2O$ dynamics also requires accurate predictions of the primary substrates (i.e. DO, $NH_4^+$, $NO_2^-$ , etc.). The experimental datasets used for calibration in lab-scale systems are either directly obtained from the reactor performance (Ding et al., 2016) or by conducting batch experiments (Ni et al., 2011). Initially, the information content of the experimental design was not studied because models aimed at describing $N_2O$ trends without focusing on rigorous calibrations (Law et al., 2011). However, the amount and quality of data of the experimental design directly impact the calibration results (Dochain and Vanrolleghem, 2001).

Some studies report the proposed calibration framework (Guo and Vanrolleghem, 2014), but the $N_2O$ parameter estimation procedures are often ill-described, with little information about each step. For example, the parameter subset selection considered during parameter estimation is sometimes not addressed. Local sensitivity measures are used as rankings for parameter selection (Pocquet et al., 2016; Spérandio et al., 2016), but these rankings are dependent on the initial parameter values and do not capture parameter interactions (Brun et al., 2001).

The overall fit and capabilities to describe $N_2O$ dynamics has relied on analysis from best-fit simulations (e.g. $R^2$), which can lead to ambiguous results that cannot discriminate between models (Lang et al., 2017; Pan et al., 2015). A more rigorous analysis of residuals (e.g. Gaussian distributions, autocorrelation functions (ACF), F-test, etc.) would benefit the validation of the model response (Bennett et al., 2013).

Also, addressing the practical identifiability of newly estimated parameters will improve $N_2O$ model discriminations procedures. For example, the parameter variance and correlation matrix are indicators of the confidence that can be given to a value, but they are not always reported, which makes it difficult to compare between $N_2O$ model predictions (Ding et al., 2016; Kim et al., 2017; Pocquet et al., 2016; Spérandio et al., 2016). Practical identifiability problems might contribute to the large variability of parameter values in $N_2O$ models (Domingo-Félez et al., 2017).

The uncertainty obtained during calibration translates into confidence intervals for model predictions. The accuracy, or width of the confidence interval, associated to the $N_2O$ emissions will be a key factor to consider during the development of mitigation strategies. Yet, the uncertainty of $N_2O$ emissions associated to model calibration is not studied.

The objective of this study is to demonstrate and evaluate a standardized procedure for parameter estimation from $N_2O$ models that relies on respirometric assays and in particular its application to analyse and validate the recently developed NDHA models. These assays are designed to allow the sequential fit of model components. The novelty resides in improving $N_2O$ calibration procedures by targeting sources of uncertainty. Subsequently, the calibration results and associated uncertainty are evaluated. The calibration approach presented is a rigorous tool beneficial for $N_2O$ model discrimination.

## 2. Material and Methods

### 2.1. Experimental Design

**Nitrifying enrichment culture.**

A lab-scale nitrifying sequencing batch reactor (5 L) was operated and displayed stable performance for three months after enrichment from an AS mixed liquor sample. Synthetic wastewater (modified after Graaf et al., 1996) with $NH_4^+$ as the only nutrient was fed at 0.5 g $NH_4^+$-N/L·d and a constant aeration rate maintained oxygen-limited conditions (DO below 0.25 mg/L) (SI-S1). $NH_4^+$ removal was 82 ± 14%, and nitritation efficiency ($NO_2^-/NH_4^+{}_{removed}$) at 85 ± 24%. The biomass composition, based on 16 rRNA targeted qPCR analysis had a dominance of AOB over NOB (30:1). Detailed information of the qPCR analysis can be found in (Terada et al., 2010).

**Monitoring nitrification and $N_2O$ production via extant respirometric assays.**

Biomass samples were harvested towards the end of the react cycle by centrifugation at 3600 g for 5 min, washed and resuspended in nitrogen-free mineral medium three times to eliminate any soluble substrate.

Assays were performed in parallel at 25°C in two 400-mL jacketed glass vessels completely filled with biomass and sealed with the insertion of a Clark-type polarographic DO electrode (YSI Model 5331, Yellow Springs, OH). Biomass samples were saturated with air or pure oxygen prior to the initiation of the respirometric assays. A decrease in the DO level in the vessel due to substrate oxidation was measured by the DO probe and continuously acquired by a personal computer interfaced to a DO monitor (YSI Model 5300, Yellow Springs, OH) by a multi-channel data acquisition device (LabPC+, National Instruments, Austin, TX). DO profiles were acquired at a user-defined frequency below the response time of the sensor (0.2 Hz). Liquid $N_2O$ concentrations were measured with Clark-type microsensors (N2O-R, Unisense A/S, Aarhus, Denmark) and pH (WTW GmbH, Weilheim, Germany). Stock solutions for all the reagents were prepared from high-purity chemicals for $NH_4HCO_3$, $NH_2OH{\cdot}HCl$, $NaNO_2$, $C_3H_5NaO_2$ (Sigma Aldrich) and by sparging ≥99.998% gas in deionized water for $N_2O$ (Sigma-Aldrich).

**Table 1** – Experimental design for respirometric assays. Grey shading corresponds to anoxic experiments.

| Scenario | Substrate added | Targeted processes | $N_2O$ pathways |
|---|---|---|---|
| Scen_AMO | $NH_4^+$ | $NH_4^+$ removal by AOB<br>$N_2O$ production at excess/limiting DO ($NH_4^+$ excess) | NN, ND |
| Scen_AMO_DO | $NH_4^+$ | Low DO part of Scen_AMO | |
| Scen_HAO | $NH_2OH$ | $NH_2OH$ removal by AOB<br>$N_2O$ production at excess/limiting DO ($NH_2OH$ excess) | NN, ND |
| Scen_NOB | $NO_2^-$ | $NO_2^-$ removal by NOB<br>$N_2O$ production at excess/limiting DO ($NO_2^-$ excess) | HD |
| Scen_An_AOB | $NH_4^+$, $NH_2OH$, $NO_2^-$ | Role of $NH_4^+$, $NH_2OH$, $NO_2^-$ on AOB-driven $N_2O$ production | NN, ND |
| Scen_An_HB | $N_2O$, $NO_2^-/NO_3^-$ | Role of $N_2O$, $NO_2^-$ and $NO_3^-$ on HB-driven $N_2O$ production | HD |

**Experimental Design**

The aim of the experimental design was to obtain informative data on $N_2O$ dynamics for a nitrifier dominated biomass to allow estimation of parameters of the NDHA model, which captures processes associated with nitrification, denitrification, and abiotic processes. Respirometric approaches were exclusively taken (on-line, high-rate $O_2$ and $N_2O$ measurements) as they allow accurate parameter estimates compared to substrate depletion experiments (Chandran et al., 2008). The kinetics of the oxidation of the primary N-substrates ($NH_4^+$, $NH_2OH$ and $NO_2^-$) were individually and step-wise, measured via extant respirometry under various initial DO conditions, with continuous $N_2O$ measurements (Table 1). Then the interaction between the different N species was ascertained. In addition, specific experiments were conducted to measure the heterotrophic and abiotic contributions to total $N_2O$ production during nitrification.

The purpose was to predict the fate of the primary N-substrates based on the specific oxygen-consuming rate. By sequentially adding substrate pulses from oxidized to reduced form ($NO_2^-$ → $NH_2OH$ → $NH_4^+$), the individual rates can be isolated. The $N_2O$ dynamics from $NH_4^+$ removal at varying $NO_2^-$ and DO concentrations can be investigated simultaneously.

### 2.2. $N_2O$ model description: NDHA

The NDHA model was proposed as a consilient model to describe $N_2O$ dynamics under a variety of conditions for biomass containing both autotrophic and heterotrophic fractions (Domingo-Félez and Smets, 2016). It considers $N_2O$ production from two autotrophic and one heterotrophic biological pathways, plus abiotic $N_2O$ formation based on recent findings (Soler-Jofra et al., 2016). Unlike any other model, NDHA can qualitatively capture NO and $N_2O$ profiles that have observed at high and low DO (Castro-Barros et al., 2016; Kampschreur et al., 2008; Rodriguez-Caballero and Pijuan, 2013; Yu et al., 2010). Here we aim to calibrate the NDHA model.

The following summarizes the essential and unique components of NDHA; for more information see (Domingo-Félez and Smets, 2016). The two autotrophic pathways have two different NO-producing processes, which are combined into a single $N_2O$-producing process. In Nitrifier Nitrification (NN), $NO_{NN}$ is produced during $NH_4^+$ oxidation (AOR) under oxic conditions. Higher AOR will likely increase $NO_{NN}$ and also $N_2O$. A fraction of $NH_4^+$, proportional to AOR is always released as $N_2O_{NN}$. In autotrophic denitrification (ND), under low DO $NO_{ND}$ is produced by the reduction of $HNO_2$ with $NH_2OH$. This step is negatively affected by DO. The reduction of both $NO_{ND}$ and $NO_{NN}$ is lumped in one process with no oxygen inhibition as it is not known whether both NIR and NOR steps are directly inhibited by DO (Kozlowski et al., 2014). Thus, if NIR is inhibited by DO the overall ND-associated $N_2O$ production will be indirectly limited. The 4-step denitrification model was considered based on (Hiatt and Grady, 2008). Individual process rates and inhibition/substrate coefficients were used as suggested for systems with low substrate accumulation. Nitrification produces $HNO_2$ and $NH_2OH$. Abiotically $NH_2OH$ can form HNO which dimerizes via $H_2N_2O_2$ to $N_2O$ and $H_2O$ (Eq. i). HNO accumulation could occur due to an imbalance between the two reactions, leading to chemical $N_2O$ production (Igarashi et al., 1997). Nitrosation of $NH_2OH$ (Eq. ii) has also been postulated as a relevant reaction in partial nitrification reactors (Soler-Jofra et al., 2016).

$$NH_2OH \rightarrow HNO \rightarrow H_2N_2O_2 \rightarrow N_2O + H_2O \qquad \text{(Eq. i)}$$

$$NH_2OH + HNO_2 \rightarrow N_2O + H_2O \qquad \text{(Eq. ii)}$$

### 2.3. Parameter estimation procedure

The steps in the parameter estimation procedure were to (1) estimate the best fit parameters to describe $O_2$ consumption and $N_2O$ production during the various (or during each type of) experimental scenarios (Figure 1), (2) estimate the contribution of separate pathways to the total $N_2O$ production, and (3) quantify the uncertainty of model predictions.

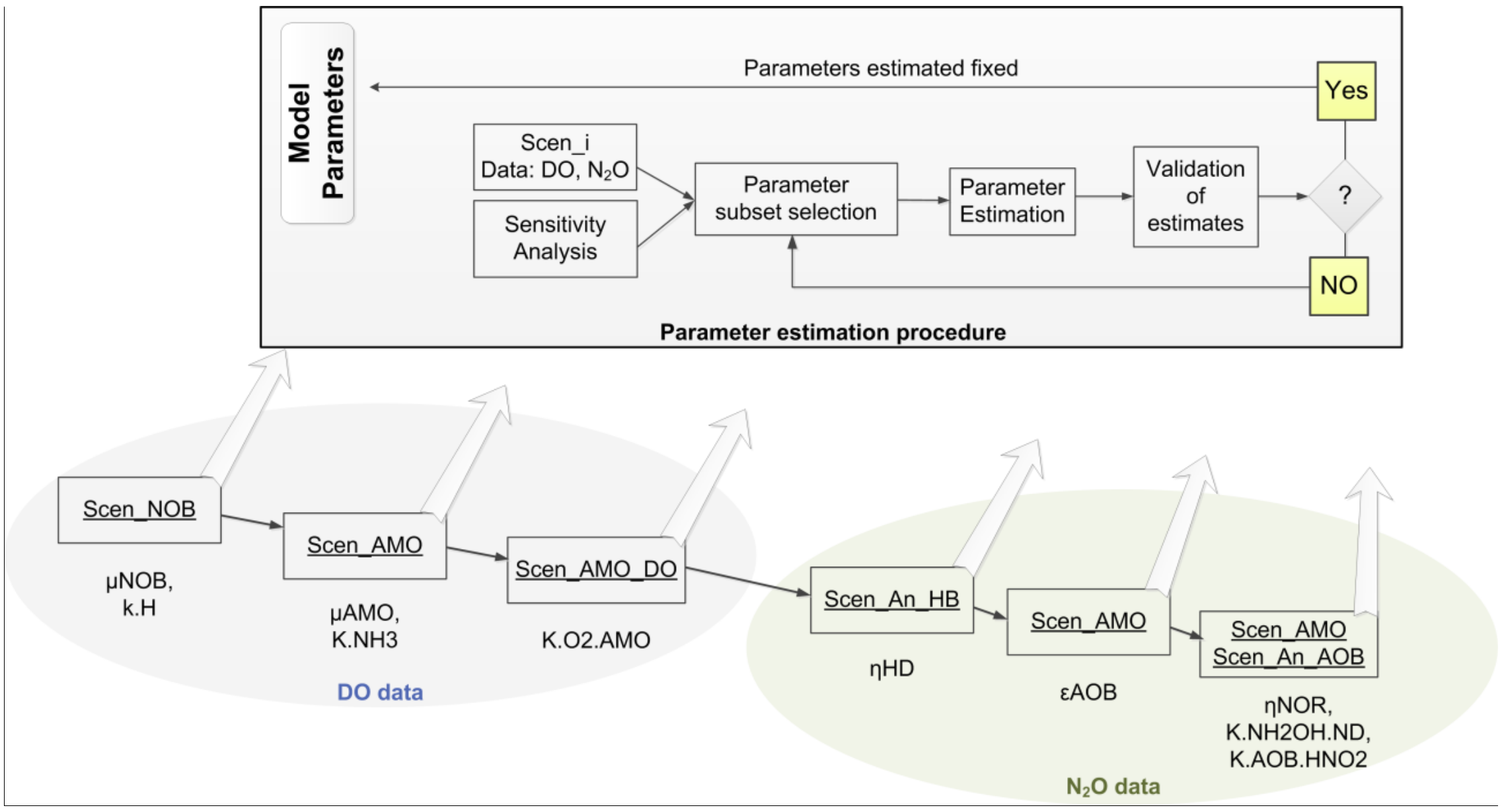

**Figure 1 –** Schematic of the parameter estimation procedure.

### 2.3.1. Scenarios (Batches grouped by substrate added: $NH_4^+$, $NH_2OH$, $NO_2^-$)

A scenario (e.g. Scen_AMO) was defined as a group of experiments with the same primary N-substrate added by pulses (Table 1). The overall oxygen consumption was the additive effect of several independent oxygen consumption processes, potentially including endogenous -, $NO_2^-$ -. $NH_2OH$ -, and $NH_4^+$-respiration. By sequentially following the respirometric response from more to less oxidized state N-substrates (i.e., first $NO_2^-$, then $NH_2OH$, then $NH_4^+$), the identifiability of nitrification kinetic parameters increases as steps can be estimated sequentially (Brouwer et al., 1998).

### 2.3.2. Parameter estimation

The error function for problem minimization was defined as:

$$\text{Error} = \sum_{j}^{m} \frac{1}{n} \sum_{i}^{n} \left( \frac{y_{sim,i} - y_{obs,i}}{\sigma_i} \right)^2$$

Where $m$ is the number of experiments in one scenario (e.g. 2 NOB experiments in Scen_NOB), $n$ the number of experimental points of each experiment, $y_{sim,i}$ the model prediction and $y_{obs,i}$ the experimental data at time $i$, and $\sigma_i$ the standard error of the experimental data. The minimization problem was started with global search method over a wide parameter space (GlobalSearch algorithm). From the estimated minimum, multiple local searches (PatternSearch algorithm) were started randomly in a narrower parameter space to avoid local minima. Model simulations were performed in the Matlab environment (The Mathworks Inc., Natick, USA).

### 2.3.3. Parameter subset selection - Global sensitivity analysis

A global sensitivity analysis was performed to identify the parameters most determinant of model outputs by linear regression of Monte-Carlo simulations (Sin et al., 2009). Uncertainty from model parameters was propagated as 10-25-50% uniform variations from their default value to model outputs (SI-S2). Latin hypercube sampling was used to cover the parameter space. The Standardized Regression Coefficient method was used to calculate the sensitivity measure $\beta_i$, which indicates the effect of the parameter on the corresponding model output (Campolongo and Saltelli, 1997) (convergence found with 500 samples).

Parameters describing the elemental biomass composition (e.g. $i_{NXB}$), yield and temperature coefficients were fixed at default values and not considered for calibration. For each scenario the top ranked most sensitive parameters were preliminary selected as candidates for parameter estimation. All possible combinations of parameter candidates were assessed by increasing the size of the calibration subset to find the largest identifiable subset with the lowest error, assessed by the Akaike Information Criterion (AIC) (Akaike, 1974). To compare the information content of different parameter subsets of the same size the optimal experimental design criteria modE (Dochain and Vanrolleghem, 2001) was calculated together with RDE, which captures the accuracy and precision of a calibrated subset (Machado et al., 2009). Newly estimated parameters were fixed at their best-fit estimate on the next calibration step (Figure 1).

### 2.3.4. Validation of model response and parameter estimates

To test the validity of the model response (i.e. the adequacy of model to predict the observed data points) the residuals ($y_{sim,i}$ - $y_{obs,i}$) were compared to a Gaussian distribution with a one-sample Kolmogorov-Smirnov test (Lilliefors, 1967). Interdependency of residuals was analysed by autocorrelation for different lag times (Cierkens et al., 2012). The quality of the model fit was calculated via correlation coefficients ($R^2$) and challenging the hypothesis of the linear regression with simultaneous unit slope and zero intercept, where a value of 0/1 indicates a bad/good model fit (F-test). Moreover, by separating the error into three components: means, slope differences and randomness the Mean Squared Error Prediction (MSEP) index identifies the main error source between randomness, mean and standard deviation of residuals (NC, ME, SE) (Haefner, 2005). The prediction accuracy and the validation of the model to individual experiments was evaluated by the Root Mean Squared Error (RMSE) and the Janus coefficient to compare the RMSE between model calibration and validation (Power, 1993).

Based on the Fisher Information Matrix (FIM) collinearity indices were calculated to evaluate the identifiability of parameter estimates (Brun et al., 2001). Approximate confidence regions were calculated following $J_{crit} = J_{opt}\left(1 + \frac{p}{N_{data}-p} F_{\alpha,p,N_{data}-p}\right)$ (Beale, 1960). Coefficients of variation (CV) were described as the ration between the variance (σ) and the mean (μ) of the estimate.

The reliability of predictive distributions (95% confidence intervals) was evaluated by calculating the Percentage of observations within the Unit Confidence Interval (PUCI) (Li et al., 2011), which combines the fraction of experimental points inside the confidence interval (PCI) and the Average Relative Interval Length (ARIL) $ARIL_{0.95} = \overline{\left(Limit_{Upper,0.95} - Limit_{Lower,0.95}\right)}/Data$ (Jin et al., 2010). A smaller ARIL value (narrow distance between upper and lower 95% CI predictions) and a larger PUCI represent a better performance.

### 2.4. Uncertainty evaluation

The effect of directly estimated versus assumed parameter uncertainty was evaluated by Monte-Carlo simulations. The parameter distribution was sampled via LHS (n = 500) for two cases: from the distributions obtained during calibration, and compared to uncertainty classes assumed from literature as a reference case (Sin et al., 2009). The resolution of prediction uncertainty was assessed by the ARIL.

## 3. Results

### 3.1. Oxygen consumption during respirometric assays.

Each scenario grouped experiments based on the substrate added: $NO_2^-$ (Scen_NOB) $NH_2OH$ (Scen_HAO) or $NH_4^+$ (Scen_AMO). In all scenarios, even prior to any substrate spikes, oxygen consumption was always positive and proportional to the biomass concentration due to endogenous respiration. After substrate addition, oxygen consumption increased, to a much higher rate with $NH_4^+$ or $NH_2OH$ than with $NO_2^-$ spikes (Figure 2). The lower $NO_2^-$ oxidizing rate of the biomass agreed with the measured low NOB vs AOB abundance (ca. 1:30).

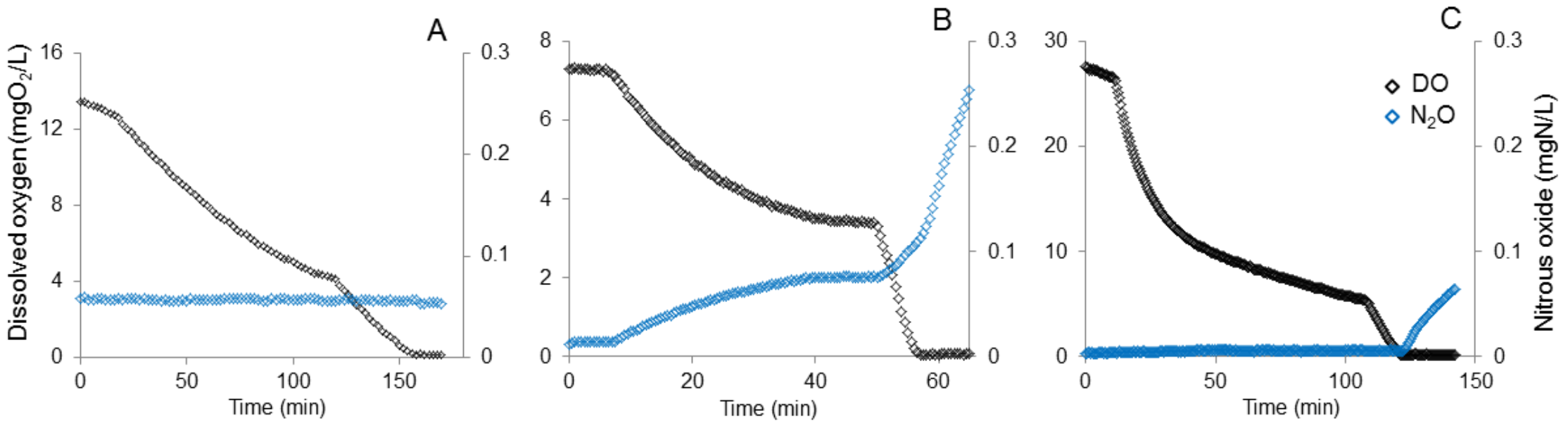

**Figure 2 –** Dissolved oxygen and liquid nitrous oxide concentrations during experiments: Scen_NOB ($NO_2^-$ pulses) (A), Scen_HAO ($NH_2OH$ pulses) (B) and Scen_AMO ($NH_4^+$ pulses) (C) (3-4.5 mgN/L).

The ranking from the global sensitivity analysis for DO shows that the sequential scenarios (measuring the respirometric response to addition of synthetic substrates) provides sufficient information to individually estimate the relevant biokinetic parameters of each step in the ammonium oxidation process (Figure 3). For example, in Scen_NOB the maximum growth rate for NOB ($\mu_{NOB}$) ranked first, while in Scen_AMO the substrate affinity ($K_{AOB.NH3}$) and maximum growth rate for the AMO ($\mu_{AOB.AMO}$) step ranked in the top.

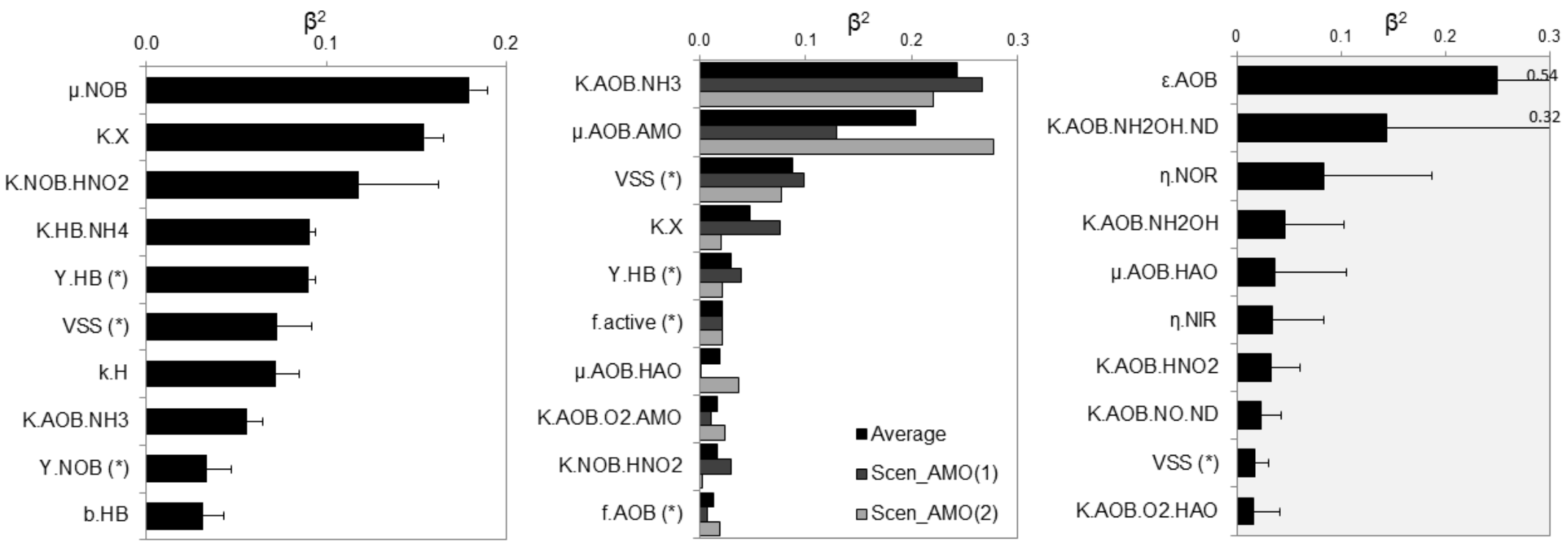

**Figure 3 –** Parameter sensitivity ranking of scenarios used during calibration for DO (white background) and $N_2O$ (grey background). Scen_NOB (left), Scen_AMO (middle and right). (*) Parameters not considered for calibration.

### 3.1.1. Estimation of primary substrate kinetics based on DO profiles

Initial conditions for each batch were defined by simulating endogenous decay and hydrolysis with default parameters. The objective of the parameter subset selection was to calibrate the minimum number of identifiable parameters that explain the data. The calibrated parameters for Scen_NOB were $\mu_{NOB}$ and $k_H$, for Scen_AMO $\mu_{AMO}$ and $K_{AOB.NH3}$, and for Scen_AMO_DO $K_{AOB.O2.AMO}$ (Table 2). To illustrate the procedure used for every scenario results from Scen_NOB and Scen_AMO are summarized in the SI (SI-S3). The electron distribution in AOB differs between $NH_4^+$ (Scen_AMO) oxidation and isolated $NH_2OH$ oxidation (Scen_HAO) (SI–S4). Hence, parameter estimation results from Scen_HAO were not considered representative of $NH_4^+$ oxidation, our targeted process. To describe more accurately the low $NH_2OH$ concentration values reported in literature (Soler-Jofra et al., 2016) the affinity for $NH_2OH$, $K_{AOB.NH2OH}$, was increased compared to other $N_2O$ models (SI–S4). The proposed higher affinity for $NH_2OH$ agrees with the lack a slower oxygen consumption rate after $NH_4^+$ depletion that would indicate $NH_2OH$ accumulation (3.43 vs 2.29 mgCOD/mgN) (Figure 2, right).

**Table 2** – Estimated parameters for each scenario (corrected for T = 20 °C). CV - coefficient of variation = variance / mean; Correlation – correlation coefficient of parameter estimates from the same scenario; RMSE – root mean squared error.

| | Scenario | Parameters | Units | Best-fit | CV | Correlation | | RMSE |
|---|---|---|---|---|---|---|---|---|
| DO data | NOB | $\mu_{NOB}$ | 1/d | 0.67 | 1.0% | 1 | -0.55 | 0.37 |
| | | $k_H$ | 1/d | 2.01 | 0.9% | -0.55 | 1 | |
| | AMO | $\mu_{AOB.AMO}$ | 1/d | 0.49 | 2.0% | 1 | 0.89 | 0.39 |
| | | $K_{AOB.NH3}$ | mgN/L | 0.12 | 3.9% | 0.89 | 1 | |
| | AMO_DO | $K_{AOB.O2.AMO}$ | mgO2/L | 0.23 | 7.0% | | | 0.08 |
| $N_2O$ data | An_HB | $\eta_{HD}$ | ( - ) | 0.055 | 0.7% | | | 0.03 |
| | AMO | $\varepsilon_{AOB}$ | ( - ) | 0.00048 | 1.1% | | | 0.001 |
| | AMO / An_AOB | $K_{AOB.HNO2}$ | µgN/L | 0.67 | 4.4% | | | 0.002 |
| | | $\eta_{NOR}$ | ( - ) | 0.16 | 3.2% | 1 | 0.98 | 0.002 |
| | | $K_{AOB.NH2OH.ND}$ | mgN/L | 0.25 | 1.8% | 0.98 | 1 | |

**Validation of model response (DO) and primary N-substrate parameter estimates**

The model consistently described the experimental DO profiles for every scenario (F-test = 1, which indicates that we fail to reject the null hypothesis of slope 1 and intercept 0 between simulations and observations) (Figure 4, A, B). The MSEP indicated that randomness was the main source of error compared to the mean or standard deviation, validating the model response during calibration (NC > ME, SE, SI-S5). The uncertainty of the parameter estimates (Table 2) was propagated to the model predictions, showing an increased resolution of the 95% predictive distributions for DO compared to the uncertainty of the reference case (ARIL = 3.8/0.5 before/after parameter estimation). The PUCI (percentage of observations bracketed by the unit confidence interval) also improved from 0.4 to 1.5.

Best-fit parameter estimates at each scenario were estimated at high accuracy: coefficients of variation (CV) were below 7% for all cases (Table 2) and the collinearity indices below 15, as suggested for identifiable subsets (Brun et al., 2002) (SI-S3). The high correlation between $\mu_{AOB.AMO}$-$K_{AOB.NH3}$ typically occurs for Monod-type kinetics but it did not affect their identifiability.

While the error distribution of each scenario was not normally distributed (Kolmogorov-Smirnov test 95%), no systematic deviations were observed (Figure 4). The analysis of the residuals indicated that for scenarios AMO and NOB the errors were autocorrelated (SI-S5). Subsequently, the effect of sampling resolution on the optimal parameter values and uncertainties was minimized until the autocorrelation obtained was negligible (SI-S5). As the sampling data frequency decreased through subsampling, the accuracy of estimates decreased too (e.g. $CV\mu_{AOB.AMO}$ = 2% point/2 min, 4.4% point/10min). However, the lower precision of the best-fit estimates did not translate into higher simulation uncertainty for the primary N-substrates ($\sigma_{95\%CI}$ increased by less than 0.06 mg/L for DO, $NH_4^+$, $NH_2OH$ or $NO_2^-$). Consequently, while the autocorrelation of residuals affected the DO parameter estimation results it did not impact the $N_2O$ calibration, the focus of this study.

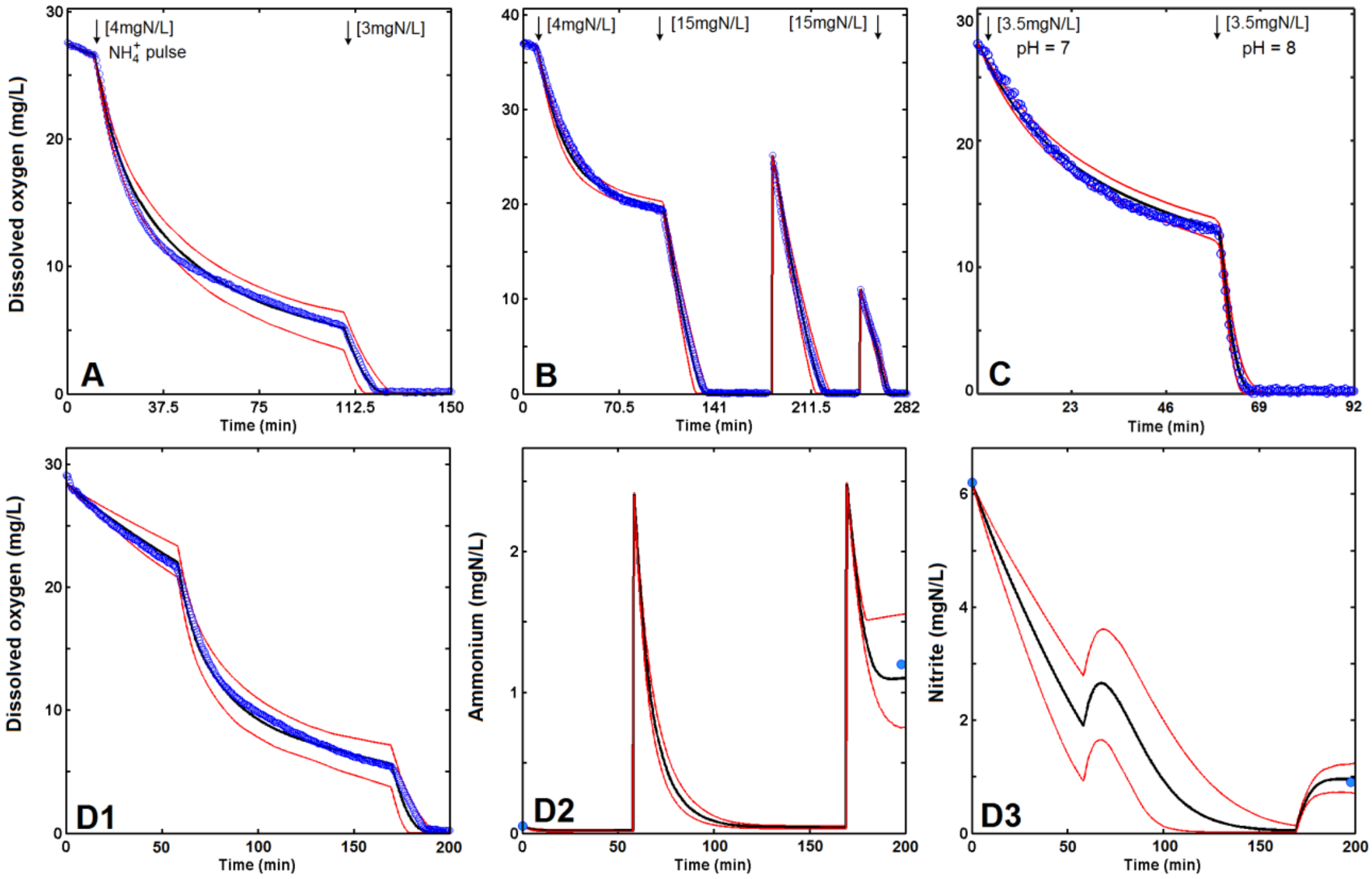

**Figure 4 –** Experimental DO, $NH_4^+$ and $NO_2^-$ (blue markers) and model predictions (black line best-fit, red lines 95% CI). Datasets from Scen_AMO, calibration: A and B; validation: C and D.

The fitted model was evaluated on five additional experiments not used during calibration with varying initial pH (7-8), $NO_2^-$ (0-6.2mgN/L) and $NH_4^+$ pulses (1-10mgN/L). The Janus coefficient and $R^2$ were close to unity (1.24 and 0.997) indicating a good model validation (Figure 4, C, D). In sum, the respirometric experimental design can be used to precisely identify and calibrate the primary substrate dynamics of the NDHA model based on the DO profiles.

### 3.2. Dynamics of $N_2O$ during different scenarios

In the same scenarios considered for DO calibration liquid $N_2O$ was also continuously measured (Figure 2). Moreover, the role of the primary N-substrates ($NH_4^+$, $NH_2OH$, $NO_2^-$, and $NO_3^-$) on $N_2O$ production was also studied under anoxic conditions (Table 1). Under conditions heterotrophic denitrification (Scen_An_HB) the presence of $NO_2^-$ and $NO_3^-$ did not show any net $N_2O$ production (data not shown). However, $N_2O$ was consumed when no other substrate was present, indicating a positive HD contribution to the total $N_2O$ pool (Figure 5).

NOB-driven $N_2O$ production was not detected during aerobic $NO_2^-$ oxidation (Figure 2, left) or at the onset of anoxia in the presence of $NO_2^-$ and $NO_3^-$.

AOB-driven $NH_4^+$ oxidation (Scen_AMO) produced a small amount of $N_2O$ under aerobic conditions and significantly increased at the onset of anoxia (Figure 2). The specific $N_2O$ production rate (mg$N_2O$-N/gVSS.min) obtained in duplicate experiments – carried out at varying biomass concentrations – were in close agreement, thus indicating biologically-driven $N_2O$ production (SI-S6).

The role of $NH_2OH$ as a direct precursor of $N_2O$ was investigated in Scen_HAO. Under aerobic conditions, $NH_2OH$ oxidation produced more $N_2O$ than $NH_4^+$ oxidation. In addition, upon reaching anoxia the $N_2O$ production rate also increased in the presence of $NH_2OH$ (Figure 2, middle). Under anoxic conditions (Scen_An_AOB) neither $NH_4^+$ nor $NO_2^-$ produced $N_2O$ individually (SI-S6). The spike of $NH_2OH$ yielded the largest amount of $N_2O$ specific to the amount of nitrogen spiked, and $NO_2^-$ was not detected at the end of the experiment ($NO_2^- < 0.05$ mgN/L). Hence, $NH_2OH$ oxidation by AOB produces $N_2O$ and does not require $O_2$. The addition of an electron acceptor like $NO_2^-$ to ongoing anoxic $NH_2OH$ oxidation increased the net $N_2O$ production rate, while addition of an electron donor as $NH_4^+$ did not (SI-S6).

Taken together, the $N_2O$ production observed in all the scenarios can only be effectively predicted using the NDHA model compared to other $N_2O$ models (Ding et al., 2016; Domingo-Félez and Smets, 2016; Ni et al., 2014; Pocquet et al., 2016).

### 3.2.1. Parameter estimation from $N_2O$ dynamics

In the absence of stripping, heterotrophic denitrification is the only $N_2O$ consuming process. First, the $N_2O$ consumption potential of the biomass was estimated and the reduction factor was considered representative for the 4-step heterotrophic denitrification processes ($\eta_{HD}$ = 0.055) (Figure 5, A). Then, $N_2O$ production observed from $NH_4^+$ oxidation at high DO, where the interference of the two denitrifying pathways is minimum, was used to calibrate the NN pathway. $\varepsilon_{AOB}$, the most sensitive parameter at high DO was calibrated, $\varepsilon_{AOB}$ = 0.00048 (Figure 5, B, SI-S2). Experiments from Scen_AMO were designed to reach anoxia at varying $HNO_2$ concentrations (0.15-3 μgN/L). Parameters associated to the ND pathway were the most sensitive and were thus calibrated ($\eta_{NOR}$ = 0.16, $K_{AOB.HNO2}$ = 0.67 μgN/L, $K_{NH2OH.ND}$ = 0.25 mgN/L) (SI-S2, S7). The abiotic contribution measured was low and not considered during parameter estimation (SI-S6).

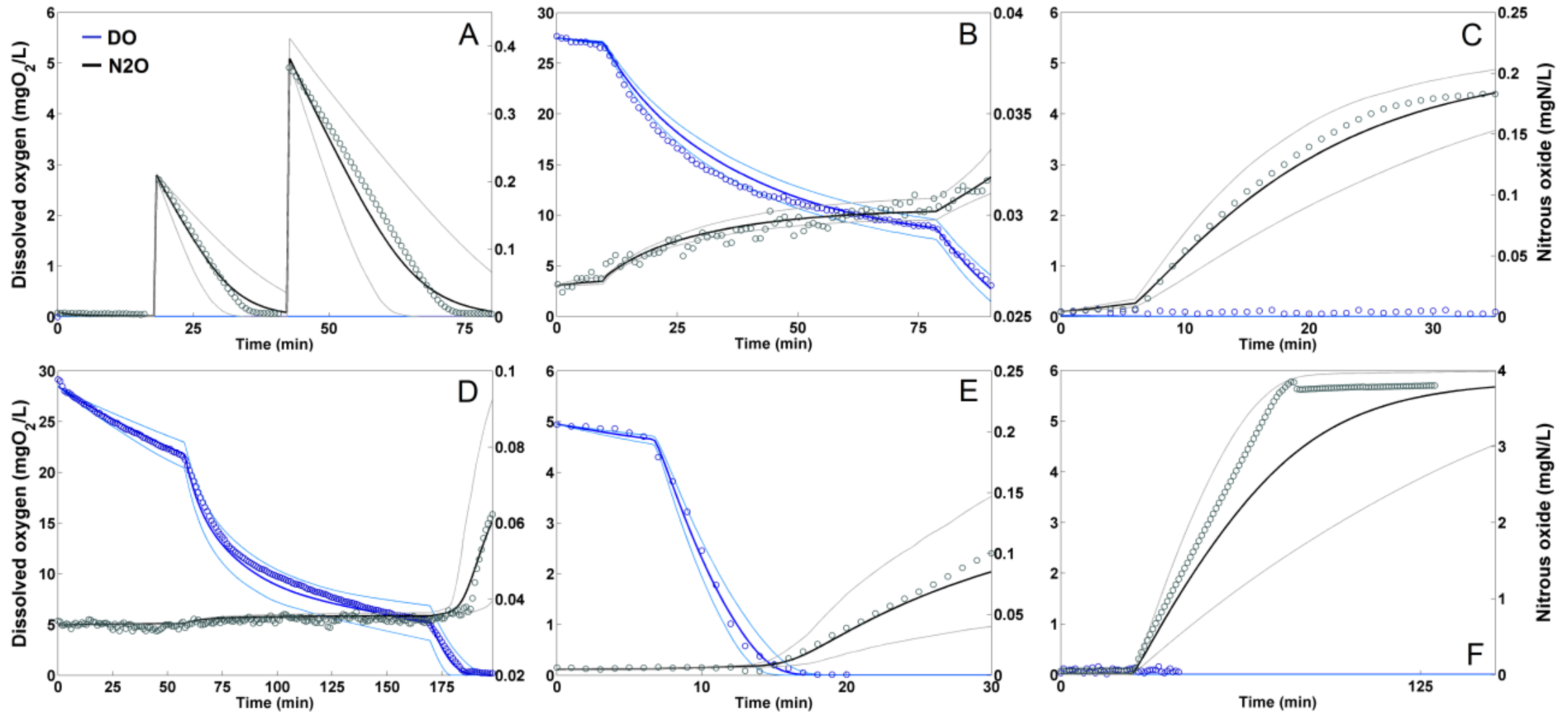

**Figure 5 –** Experimental $N_2O$ (blue markers) and model predictions (dark line best-fit, light lines 95% CI) for the $N_2O$ calibration. Datasets from calibration: (A) Scen_An_HB ($N_2O$ pulse), (B) Scen_AMO (Aerobic $NH_4^+$ pulse), (C) Scen_An_AOB (Anoxic $NH_2OH$ pulse). Datasets from validation: (D, E) Scen_AMO (Aerobic → anoxic $NH_4^+$ pulse), (F) Scen_An_AOB ($NH_2OH$ pulse).

**Validation of model response and secondary substrate ($N_2O$) parameter estimates**

The calibrated NDHA model described the $N_2O$ production dynamics and yield observed in the calibration datasets (F-test = 1). In all but one of the assays randomness was the most important part of the error based on the MSEP analysis (SI-S5). After calibration the ARIL narrowed by 58% from the original resolution and the PUCI increased by 71% (n = 6 assays).

The predictive ability of the model was evaluated on three batches with lower $HNO_2$ and with higher $NH_2OH$ pulses ($HNO_2$ < 0.15 μgN/L, $NH_2OH$ = 2 mgN/L). The average Janus coefficient of the validation prediction was 1.57 and $R^2$ was 0.985, indicating a good validation (Figure 5, D, E, F). Hence, the NDHA model could describe the $N_2O$ production rates at a range of DO and $HNO_2$ concentrations.

The simple experimental design allowed the isolation of the various components of $N_2O$ dynamics during $NH_4^+$ oxidation, and the parameter estimation procedure the identification of relevant model parameters.

### 3.3. Model predictions under varying DO and $HNO_2$: Scenario analysis

To investigate the effect of DO and $HNO_2$ on $N_2O$ production the NDHA model was evaluated at varying DO and $HNO_2$ concentrations at pH = 7.5 (Figure 6) with the newly estimated parameters. The model predicted the largest $N_2O$ emission at the lowest DO and high $HNO_2$ (> 20%, SI-S8); and the lowest $N_2O$ emission at the highest DO and lowest $HNO_2$ (0.13%). The effect of increasing $HNO_2$ is seen at every DO level ($DO_{0.3}$: 0.33 → 5.4%, $DO_{5.0}$: 0.13 → 0.28%). Conversely, increasing DO lowered the $N_2O$ emission factor regardless of the $HNO_2$ level. The NO emission showed an increasing pattern with $HNO_2$ but a minimum was found at DO = 2.0 mg/L, further increasing at higher DO (SI-S8).

The contribution of the NN pathway was maximum when $HNO_2$ was not present and decreased with increasing $HNO_2$, at a faster rate at lower than at higher DO (2.4 and 47% respectively). The ND contribution followed opposite trends, indicating a shift between autotrophic pathways driven by $HNO_2$ and DO. The ND contribution increased with $HNO_2$, at a steeper rate at lower DO (97.4%) than at higher (53%). The HD contribution was maximum at low DO and high $HNO_2$ but only reached 0.2% (SI-S8).

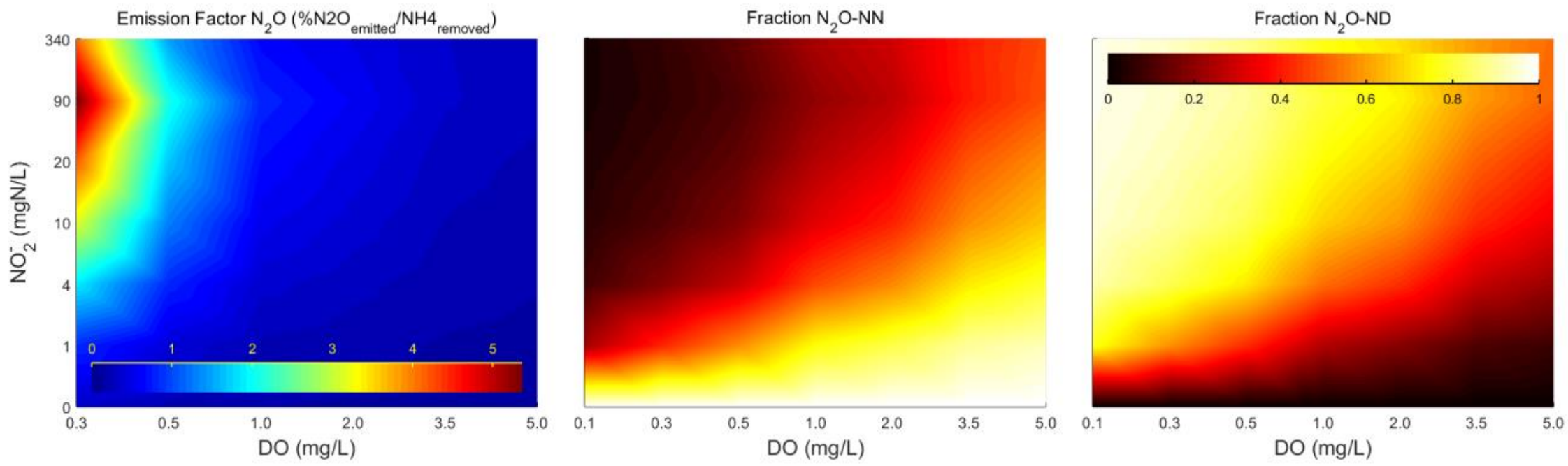

**Figure 6 –** Scenario analysis using the validated NHDA simulation model. Simulations were run under constant DO levels (0.1 - 0.3 - 0.5 - 1.0 - 2.0 - 3.5 - 5.0 mg/L), $NO_2^-$ (0 – 1 – 4 – 10 – 20 – 90 – 340 mgN/L). (Left) $N_2O$ emission factor (% $N_2O/NH_4^+$), colorbar: 0 – 5%, blue - red. (Middle, Right) NN, ND Pathway contribution ( - ), colorbar: 0 – 1, black - white.

## 4. Discussion

### 4.1. Parameter estimation from respirometric assays: oxygen consumption

The respirometric experiments were used to investigate the oxygen-consuming processes driven by the AOB-enriched biomass in the presence of reduced N-species ($NH_4^+$, $NH_2OH$ and $NO_2^-$). If a model captures accurately the relevant oxygen-consuming processes, then DO and the primary N-substrates are predicted accurately. The experimental design based on the concatenated oxygen consumption allowed the isolation of individual processes independently (endogenous → $NO_2^-$ → $NH_4^+$) (Chandran and Smets, 2005).

The calibrated model could describe the endogenous oxygen uptake and $NO_2^-$ oxidation in Scen_NOB. However, because of the low NOB abundance the oxygen consumption from $NO_2^-$ oxidation was low, shown by a similar sensitivity of NOB and endogenous parameters to oxygen consumption after $NO_2^-$ spikes (Figure 3).

In Scen_AMO oxygen consumption was very sensitive to $NH_4^+$ dynamics (SI-S2), which yielded precise estimates for $\mu_{AOB.AMO}$, $K_{AOB.NH3}$ and $K_{AOB.O2.AMO}$ (Table 2). The maximum AOB growth rate ($\mu_{AOB.AMO}$ = 0.49 1/d) is in the low range of literature values found for N. *europaea* (0.56-1.62 1/d) (Brockmann et al., 2008). The biomass concentration ($X_{AOB}$), growth yield ($Y_{AOB}$) and maximum growth rate cannot be simultaneously identified from short experiments solely with DO data (Ellis et al., 1996; Petersen et al., 2001). Hence, the estimated growth rate is linearly dependent on the fixed values for $X_{AOB}$ and $Y_{AOB}$: a lower initial condition for $X_{AOB}$ would yield a higher estimate for $\mu_{AOB.AMO}$. Overall, the maximum specific $NH_4^+$ oxidation, 7.54±0.1E-05 gN/gVSS$_{(AOB)}$/h, was similar to other literature values for an AOB-enriched biomass (Ciudad et al., 2006). For the same $NH_4^+$ concentration, the higher oxygen consumption rate observed at higher pH was predicted by considering $NH_3$ the true substrate. The estimated affinities for both $NH_4^+$ oxidation substrates ($K_{AOB.NH3}$ = 0.12 mgN/L, $K_{AOB.O2.AMO}$ = 0.23 mg/L) were in range of literature values (Magrí et al., 2007; Park and Noguera, 2007).

Overall, the precision of the identified parameter was high (CV < 7%), common from respirometric studies (Petersen et al., 2001). It should be noted that the concentration of the spikes did not include uncertainty and was not estimated, which decreased the uncertainty of model predictions (Gernaey et al., 2002).

### 4.2. Role of $NH_4^+$ oxidation intermediates on $N_2O$ production: experimental and modelling results

Nitrification plays an important role on $N_2O$ emissions from N-removing systems, where $NH_4^+$, $NO_2^-$ and DO are the main substrates. Experimental results indicated that aerobic $NH_4^+$-oxidation products, $NH_2OH$ and $NO_2^-$ were responsible for the higher $N_2O$ production rate at the onset of anoxia and not $NH_4^+$ itself, which requires molecular $O_2$ for its oxidation (Sayavedra-Soto et al., 1996) (SI-S6). $NH_2OH$ has been shown to be a key compound regulating $N_2O$ production by AOB (Caranto et al., 2016; de Bruijn et al., 1995; Kozlowski et al., 2016). Because of its high reactivity under aerobic and anoxic conditions, it is an important electron donor for the cytochrome pool of AOB. Previous studies have shown the higher $N_2O$ yield of nitrifying biomass and pure cultures fed on $NH_2OH$ compared to $NH_4^+$, also observed in Scen_HAO ($N_2O_R_{NH2OH}/N_2O_R_{NH4+}$ = 40) (Figure 2) (Kim et al., 2010; Kozlowski et al., 2016). Here we show that even under anoxic conditions the sole presence of $NH_2OH$ also yields a large amount of $N_2O$ (SI-S6), recently suggested as a new $N_2O$ producing pathway by (cyt) P460 (Caranto et al., 2016). The addition of an electron donor like $NO_2^-$ further increased $N_2O$ production, highlighting the role of the primary N-substrates on $N_2O$ dynamics, especially of $NH_2OH$.

The NDHA model captures the observed anoxic $NH_2OH$ oxidation to $N_2O$ with no $HNO_2$ production associated. A $NH_2OH$ pulse in the concentration range of reported measurements (0.1mgN/L) could be described in the calibration dataset (Figure 5, F-test = 1, $R^2$ > 0.99); and at higher $NH_2OH$ concentrations (2 mgN/L) the model predicted the $N_2O$ trend but not as accurately (Figure 5, F-test = 0, $R^2$ = 0.97). Based on the model structure of other two-pathway models for AOB none can predict the observed $N_2O$ dynamics. In certain models $NH_2OH$ does not react under anoxic conditions (Ding et al., 2016; Pocquet et al., 2016), or reacts producing both $N_2O$ and $HNO_2$ (Ni et al., 2014).

Under a variety of DO, $HNO_2$ and $NH_3$ concentrations the calibrated NDHA model could describe the observed $N_2O$ dynamics. Other models, with varying degrees of complexity, have also described the effect of $HNO_2$ and DO (4-6 processes) but the effect of $NH_2OH$, the main driver of $N_2O$ production, was not considered (Ding et al., 2016; Ni et al., 2014). The scenario analysis indicated a shift between the main pathway contributions governed by DO and $HNO_2$ (Figure 6). This relationship has been described by other two-pathway models, where ND was the main contributor to the $N_2O$ emission factor during $NH_4^+$ oxidation and the highest $N_2O$ emission factor was observed at low DO (Ni et al., 2014; Pocquet et al., 2016).

### 4.3. $N_2O$ model calibration

#### 4.3.1. Analysis of the NDHA parameter estimates.

In the last years new $N_2O$ models have improved their best-fit predictions under different scenarios (i.e. varying DO, $NO_2^-$) by increasing the number of processes and variables considered. For example, all the models describe the ND pathway with a $NO_2^-$ dependency (Ni et al., 2011; Pocquet et al., 2016; Schreiber et al., 2009), or the NN pathway as a fraction of the $NH_2OH$ oxidation to $NO_2^-$ (Ni et al., 2014; Pocquet et al., 2016). While the intermediates of $NH_4^+$ oxidation or the process rates are described differently, some parameters are common across $N_2O$ models.

In this study, the contribution of the NN pathway ($\varepsilon_{AOB}$ = 0.048% $\mu_{AOB.HAO}$) is in the low range of other reported values (0.052-0.15%), while the maximum $N_2O$ production rate, described by $\eta_{NOR}$ = 0.16, lies in the range (0.07 – 0.34). In agreement with the ND description of the model by Pocquet et al., (2016), the electron acceptor of the ND pathway was $HNO_2$ instead of $NO_2^-$. Increasing $N_2O$ production rates were observed at higher $HNO_2$ but constant $NO_2^-$ (9.5-10mgN/L, 0.8-1.5 µgHNO2-N/L). The affinity for $HNO_2$ ($K_{AOB.HNO2}$ = 0.67 µgHNO2-N/L, 17.1 mg $NO_2^-$-N/L, pH 7.5, 20 ºC) could be estimated from experiments run at varying $HNO_2$ (0.16-1.5 µgHNO2-N/L). The affinity for $NO_2^-$ is 100 times lower than other nitrifying systems, but 15 times higher than a $NO_2^-$-accumulating biomass ($K_{AOB.NO2-}$ = 0.14, 282 mgN/L)(Pocquet et al., 2016; Schreiber et al., 2009). The difference could be explained by the operating conditions at which each biomass is acclimated: low $NO_2^-$ for activated sludge systems (≈ 0.5mgN/ L) and high $NO_2^-$ for nitritating reactors (50-150 mg $NO_2^-$-N/L in the parent reactor of this study). The NDHA model combined with the experimental design allows the simultaneous estimation of parameters describing main N-substrates and $N_2O$ dynamics from simple respirometric experiments.

#### 4.3.2. $N_2O$ models: response validation and identifiability

As $N_2O$ models produce better fits discrimination tools become more important. If the capabilities of two models to describe dynamic $N_2O$ trends are similar (Lang et al., 2017; Pan et al., 2015) visual inspection or metrics such as $R^2$ are not sufficient, and more rigorous statistics as the F-test used in this study are necessary for model discrimination.

The parameter subset selection during calibration of each scenario was based on the lower AIC criteria. The identifiability of the estimated parameters was assessed by the correlation matrix and the precision of the estimated parameters (CV < 5%). The triplet $K_{AOB.HNO2}$, $K_{AOB.NH2OH.ND}$ and $\eta_{NOR}$ were estimated with the same dataset, and based on their collinearity index ($\gamma$ > 15) they are not identifiable and their values depend on the others. This metric is based on local sensitivities, and as shown in (Table 2), the high correlation between $K_{AOB.NH2OH.ND}$ and $\eta_{NOR}$ could be responsible for the high collinearity of the triplet. An improved experimental design or an additional dataset such as NO would improve the identifiability these parameters, as shown for other two-pathway $N_2O$ models (Pocquet et al., 2016). Together with the best-fit prediction the 95% CI of the calibrated NDHA model bracketed the experimental datasets, validating the model response.

The uncertainty of estimates (CV, correlation matrix) is not always reported in literature (Ding et al., 2016; Kim et al., 2017; Spérandio et al., 2016), which hampers critical discrimination procedures. While some models have reported the estimated variance identifiability metrics are scarce and not assessed. To the authors knowledge none of the proposed $N_2O$ models has studied how the uncertainty of parameter estimates affects $N_2O$ predictions (e.g. ARIL, PUCI). Moreover, the analysis of residuals shows that high frequency data such as online sensors can lead to autocorrelated residuals in $N_2O$ measurements (SI-S5). While very high precision of estimates is possible, testing the model response can avoid a possible over interpretation of the dataset and uncertainty underestimation (CV $\ll$ 0.001% (Peng et al., 2015)). In this study we show that addressing parameter identifiability after model calibration will benefit $N_2O$ model discrimination studies.

### 4.4. $N_2O$ model uncertainty

With the final objective of designing $N_2O$ mitigation strategies, the confidence of model predictions is critical when quantifying $N_2O$ emissions. As a by-product of $NH_4^+$ oxidation, the uncertainty associated to $NH_4^+$ removal processes will propagate to $N_2O$ predictions. The respirometric experimental design allowed for accurate estimates and narrow 95% confidence intervals for DO, $NH_4^+$ and other N-species, which was critical to reduce the predicted uncertainty for $N_2O$ (SI-S9). $N_2O$ models have been calibrated by sequentially fitting the primary N-substrates followed by the $N_2O$ dynamics (Ni et al., 2014; Pocquet et al., 2016). The effect of propagating the uncertainty from the calibration of primary N-substrates to $N_2O$ was not discussed. Consequently, the precision of the $N_2O$ calibration could be overestimated.

To evaluate the performance of the parameter estimation results $N_2O$ emissions from a simulated case study were examined via Monte-Carlo simulations (pH = 7.5, $NH_4^+$ = 70 mgN/L, DO = 0.3, 1.3 mg/L and $NO_2^-$ = 0, 1, 5, 15, 100 mgN/L, 500 simulations). At pseudo-steady state the standard error of the Monte-Carlo simulations shows the propagated uncertainty. For the 10 parameters estimated in this study the normalized uncertainty associated to their default class (Sin et al., 2009) is approximately 40% of the $N_2O$ emission factor (Figure 7). If the calibration results from this experimental design are considered instead (Table 2), the uncertainty decreases to 10% (Figure 7). These results highlight the importance of considering uncertainty propagation in $N_2O$ predictions, as $N_2O$ emissions are greatly affected by the uncertainty of primary N-substrates.

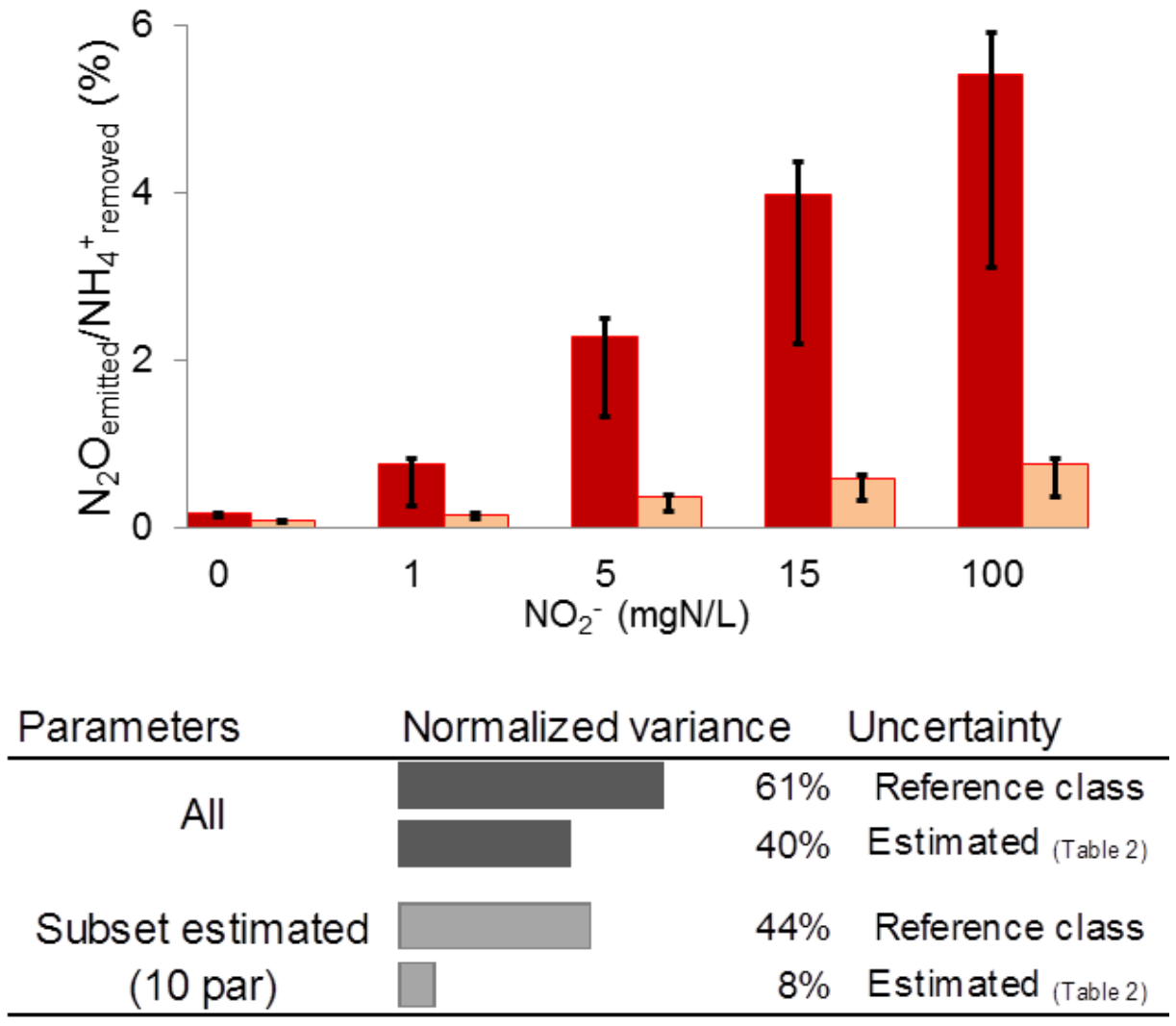

**Figure 7 –** Model evaluation results (500 runs) at pseudo-steady state for $NH_4^+$ removal at constant DO and $NO_2^-$: AOB-enriched biomass, pH = 7.5, $NH_4^+$ = 70 mgN/L, DO ([0.3 – 1.3] mg/L) and $NO_2^-$ [0 – 1 – 5 – 15 – 100] mgN/L. (Top) $N_2O$ emission factor at low DO (red) and high DO (light red); top standard error corresponds to uncertainty of estimated parameters only (Table 2), bottom standard error corresponds to uncertainty in All model parameters. (Bottom) Normalized variance for uncertainty considered in All model parameters (dark grey) or only for the 10 parameters estimated in this study (light grey) (Table 2) with default uncertainty (top bar) or from this study (Table 2) (bottom bar).

## Conclusions

- A novel experimental design to calibrate $N_2O$ models through extant respirometry is proposed that combines DO and $N_2O$ measurements.
- Parameters associated to $NO_2^-$ and $NH_4^+$ oxidation were sequentially fitted to DO consumption profiles by isolating individual processes. Five parameters were identified from the DO dataset and another five were estimated from the $N_2O$ dataset with low uncertainty (CV < 10%).
- The NDHA model response was validated and described AOB-driven $N_2O$ production at varying DO and $HNO_2$ concentrations.
- For the first time the uncertainty of the calibrated parameters was propagated to the model outputs in a simulation case study, and compared to the uncertainty from a reference case. The uncertainty of the $N_2O$ emission factor predicted was reduced from ~ 40% of its value to ~ 10%.

**Software availability**

The MATLAB/SIMULINK code containing the implementation of the model is free upon request to the corresponding author.

**Acknowledgements**

This research was funded by the Danish Agency for Science, Technology and Innovation through the Research Project LaGas (12-132633). The authors have no conflict of interest to declare.

# Supplementary Information

**Title**

Calibration of the NDHA model to describe N2O dynamics from respirometric assays.

**Author list**

Carlos Domingo-Félez[a], María Calderó-Pascual[a], Gürkan Sin[b], Benedek G. Plósz[a,1], Barth F. Smets[a]*

[a]Department of Environmental Engineering, Technical University of Denmark, Miljøvej 115, 2800 Kgs. Lyngby, Denmark

[b]Department of Chemical and Biochemical Engineering, Technical University of Denmark, Søltofts Plads 229, 2800 Kgs. Lyngby, Denmark

[1]Present address: Department of Chemical Engineering, University of Bath, Bath, England.

* Corresponding author:

Barth F. Smets, Phone: +45 4525 1600, Fax: +45 4593 2850, E-mail: bfsm@env.dtu.dk

---

## Section S1 - Parent reactor operation

During one cycle (360min) the parent nitritating reactor was fed five times (1 min) 0.5 L during the reacting aerated phase (320 min), followed by a settling (30 min), decanting (5 min) and idle phase (5 min) (HRT = 12 hours). The ammonium load was approximately 0.5 gN/L.d with an average removal efficiency of 82% (ammonium removed / loaded) and nitritating efficiency of 85% (nitrite accumulated / ammonium removed).

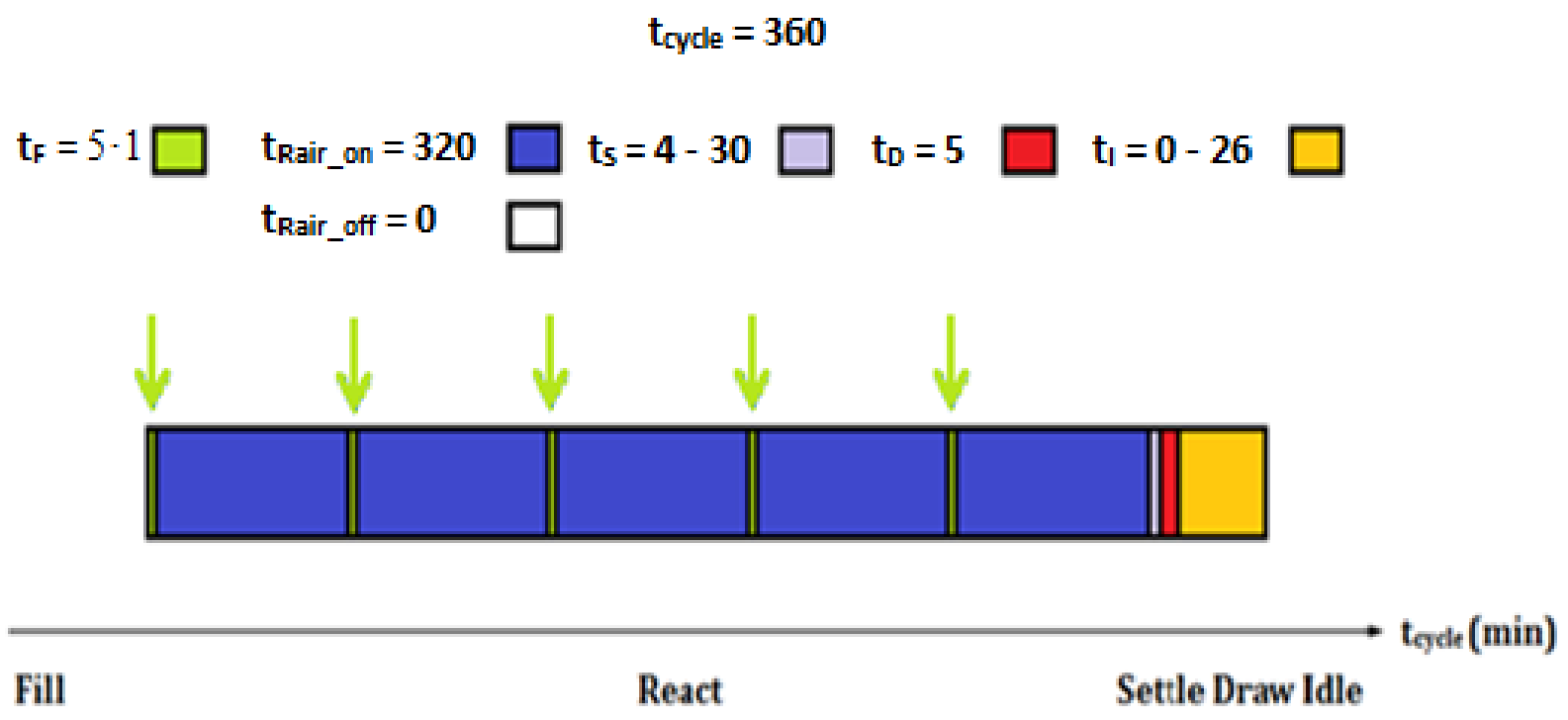

**Figure S1-1** – Operation regime for the parent nitritating SBR: 5 feeds / cycle (green), aerated during 320 minutes (blue), settling (pink), decanting (red) and idle phase (yellow).

## Section S2 – Sensitivity analysis

Dynamic cumulative $\beta^2$ values for the global sensitivity analysis were calculated for every experiment (SRC method). The uncertainty was defined by classes for each model parameter following (Sin et al., 2009) (Table S2-1). Process rate equations and stoichiometry can be found in S10.

**Table S2-1** – Uncertainty class definition for model parameters.

| Parameter ($\theta_i$) | Uncertainty Class |
|---|---|
| $i_{NXB}$, $i_{NXI}$, $i_{NXS}$, VSS, $Y_{AOB}$, $Y_{HB}$, $Y_{NOB}$, $K_La_{N2O}$, $K_La_{O2}$, $K_La_{NO}$, $f_{active}$, $f_{AOB}$, $f_{NOB}$, $f_{HB}$, $f_I$, | 1 (± 10%) |
| $b_{AOB}$, $b_{HB}$, $b_{NOB}$, $k_H$, $\eta_{HD}$, $\mu_{AOB_AMO}$, $\mu_{AOB_HAO}$, $\mu_{HB}$, $\mu_{HB_NAR}$, $\mu_{HB_NIR}$, $\mu_{HB_NOR}$, $\mu_{HB_NOS}$, $\mu_{NOB}$, $\eta_b$, $\eta_{h_anox}$, $\eta_{h_anaer}$, | 2 (± 25%) |
| $K_{AOB_NH2OH}$, $K_{AOB_NH2OH_ND}$, $K_{AOB_NH3}$, $K_{AOB_NO}$, $K_{AOB_HNO2}$, $K_{AOB_O2_AMO}$, $K_{AOB_O2_HAO}$, $K_{AOB_O2_i}$, $K_{HB_NH4}$, $K_{HB_N2O}$, $K_{HB_NO}$, $K_{HB_NO2}$, $K_{HB_NO3}$, $K_{HB_O2}$, $K_{HB_O2_i_NAR}$, $K_{HB_O2_i_NIR}$, $K_{HB_O2_i_NOR}$, $K_{HB_O2_i_NOS}$, $K_{HB_S}$, $K_{HB_S_NAR}$, $K_{HB_S_NIR}$, $K_{HB_S_NOR}$, $K_{HB_S_NOS}$, $K_{NOB_HNO2}$, $K_{NOB_O2}$, $K_{NOB_i_HNO2}$, $K_{NOB_i_NH3}$, $K_{AOB_i_HNO2}$, $K_{AOB_i_NH3}$, $K_{b_O2}$, $K_X$, $\varepsilon_{AOB}$, $\eta_{NIR}$, $\eta_{NOR}$, | 3 (± 50%) |

The regions of the dynamic SA considered where those with good regression ($R^2 > 0.7$). In specific parts of the experiment, for example during anoxic periods, the SRC method is not valid for the DO model output ($R^2 < 0.7$). These regions were thus not used during the analysis.

The empty region between the dynamic cumulative values and $\beta^2 = 1$ corresponds to the rest of parameters with very low sensitivity ($\beta^2 < 0.02$ not shown) and other parameters not considered for the analysis, and calibration: $Y_{XB}$, VSS, biomass_active_fraction, etc.

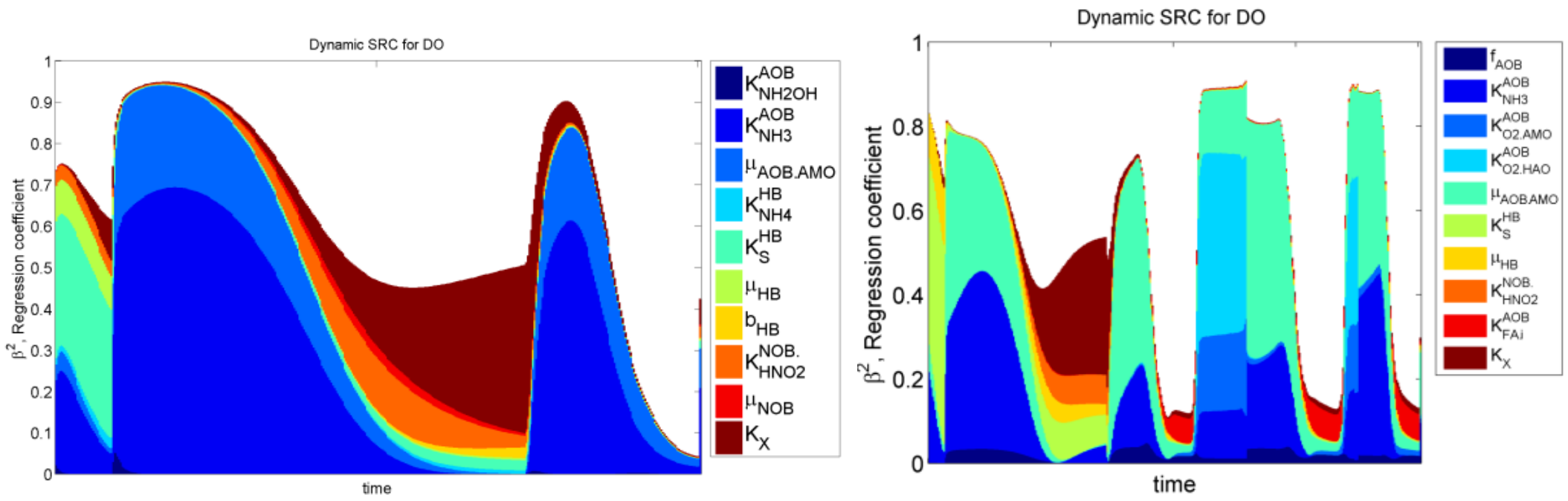

**Figure S2-1 –** Dynamic sensitivity analysis for DO: Scen_AMO ($NH_4^+$ pulses). Only parameters with squared beta ($\beta^2$) larger than 0.02 at some point of the batch are shown.

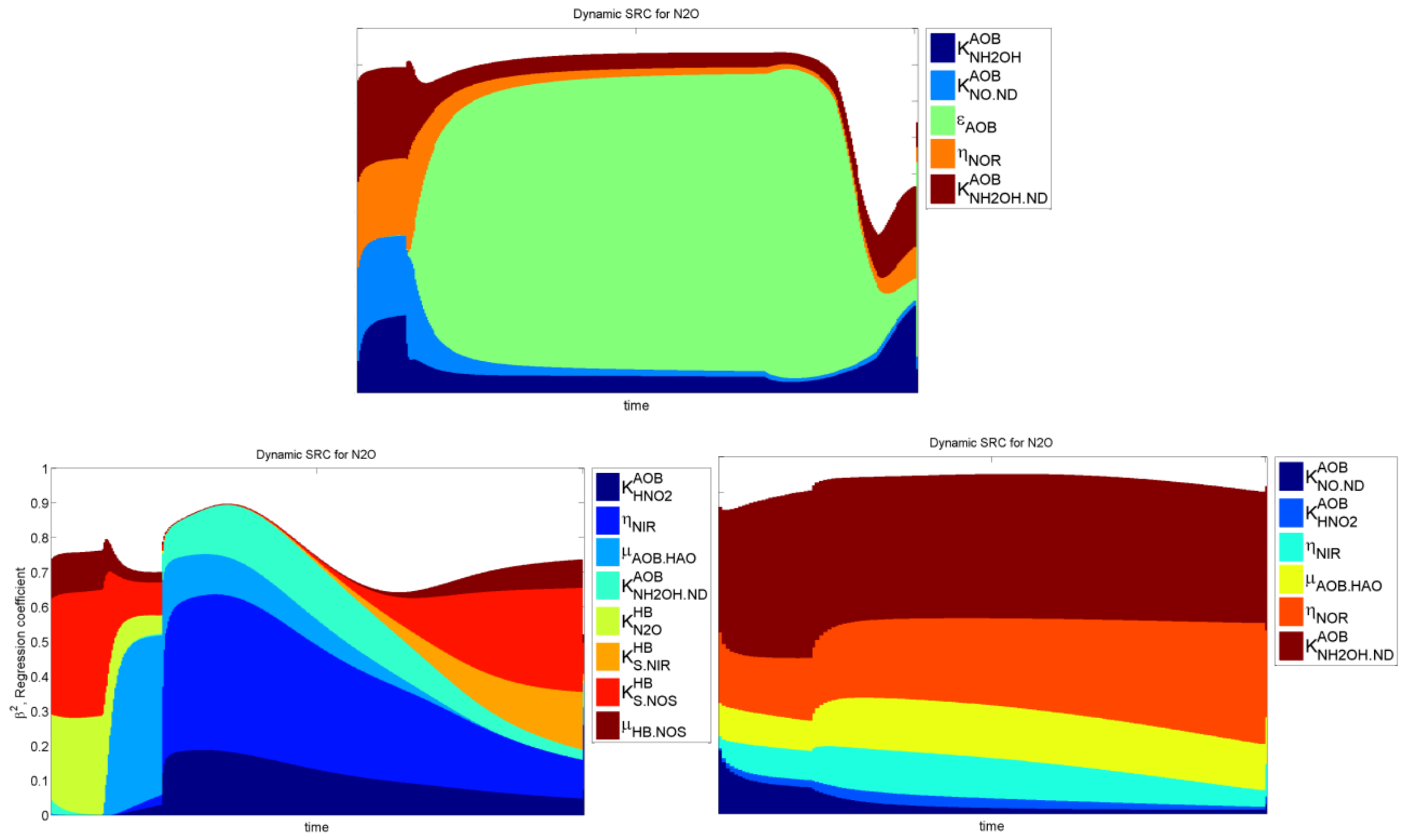

**Figure S2-2 –** Dynamic sensitivity analysis for $N_2O$: (Top) Scen_AMO ($NH_4^+$ pulses). (Bottom) Scen_An_AOB (left: $NH_2OH$ followed by $NO_2^-$ pulse; right: $NH_2OH$ pulse). Only parameters with squared beta ($\beta^2$) larger than 0.02 at some point of the batch are shown.

## Section S3 - Parameter subset selection during calibration

The objective of the calibration procedure was to find the minimum number of parameters that explain the scenario data with the less possible correlation amongst the subset. The same methodology was used for every scenario.

The AIC criterion aims to find the simplest model that describes the data, thus it is useful to compare subsets of different size (Akaike, 1974). The larger the RDE metric the better the capacity of a subset to explain the data with a low uncertainty in the estimates (Machado et al., 2009). The modE criterion describes the shape (relationship between largest/smallest axes) of the confidence region to prevent correlated parameters (Dochain and Vanrolleghem, 2001).

For a given size (n) all the subsets were evaluated. The best subset of size (n) was selected by the lowest AIC (blue) and largest RDE and 1/modE. Then, the subset size was increased (n+1) and all the possible subsets evaluated again. If the AIC value did not decrease for any of the new subsets the calibration was stopped in the subset with lowest AIC value (and largest RDE, 1/modE) of size (n).

**Table S3-1** – Parameter subset selection procedure for Scen_NOB.

| Parameters | AIC | log(RDE) | 1/modE | Optimized values | Corrrelation | Collinearity |
|---|---|---|---|---|---|---|
| µ.NOB | | | | 0.00072 | | |
| K.X | | | | 0.17 | | |
| K.NOB.HNO2 | | | | 0.00013 | | |
| K.HB.NH4 | | | | 0.013 | | |
| k.H | | | | 0.0017 | | |
| µ.NOB / K.X | | | | 0.0008 - 0.17 | 0.58 | 3.3 - 1.4 |
| µ.NOB / K.NOB.HNO2 | | | | 0.0023 - 0.00088 | 0.99 | 35 - 24 |
| µ.NOB / K.HB.NH4 | | | | 0.0008 - 0.013 | 0.99 | 31 - 11 |
| µ.NOB / k.H * | | | | 0.00078 - 0.0017 | -0.58 | 3.2 - 1.4 |
| K.X / K.NOB.HNO2 | | | | 0.17 - 0.0001 | -0.56 | 4.1 - 1.5 |
| K.X / K.HB.NH4 | | | | 0.18 - 0.0029 | -0.34 | 3.1 - 1.5 |
| K.X / k.H | | | | 0.22 - 0.0022 | 0.96 | 148 - 1220 |
| K.NOB.HNO2 / K.HB.NH4 | | | | 0.000019 - 0.025 | -0.97 | 6.1 - 8.4 |
| K.NOB.HNO2 / k.H | | | | 0.00010 - 0.0017 | 0.63 | 4.1 - 1.5 |
| K.HB.NH4 / k.H | | | | 0.010 - 0.0017 | 0.61 | 3.5 - 1.5 |

(*) Parameter subset selected for calibration. Maximum rates are shown as 1/minutes.

**Table S3-2** – Parameter subset selection procedure for Scen_AMO.

| Parameters | AIC | log(RDE) | 1/modE | Optimized values | Corrrelation | Collinearity |
|---|---|---|---|---|---|---|
| K.NH3 | | | | 0.25 | | |
| K.O2.AMO | | | | 3.02 | | |
| µ.AMO | | | | 0.0044 | | |
| µ.HAO | | | | 0.00035 | | |
| K.X | | | | 0.173 | | |
| K.NH3 / K.O2.AMO | | | | 0.18 - 3.8 | 0.30 | 3.2-3.3 |
| K.NH3 / µ.AMO * | | | | 0.12 - 0.0033 | 0.89 | 22.4-5.2 |
| K.NH3 / µ.HAO | | | | 0.19 - 0.00034 | -0.65 | 1.8-2.1 |
| K.NH3 / K.X | | | | 0.27 - 0.098 | 0.76 | 1.9-1.9 |
| K.O2.AMO / µ.AMO | | | | 1.6 - 0.0048 | -0.91 | 2.5-3.7 |
| K.O2.AMO / µ.HAO | | | | 2.0 - 0.00041 | 0.71 | 1.4-1.5 |
| K.O2.AMO / K.X | | | | 3.5 - 0.10 | -0.02 | 2.2-1.5 |
| µ.AMO / µ.HAO | | | | 0.0044 - 0.0018 | -0.81 | 2.8 - 1.0 |
| µ.AMO / K.X | | | | 0.0043 - 0.11 | -0.75 | 1.7-1.8 |
| µ.HAO / K.X | | | | 0.00035 - 0.14 | -0.06 | 1.1-1.3 |
| K.NH3 / µ.AMO / K.O2.AMO | | | | 0.10 - 0.0031 - 0.55 | 0.04/-0.95/0.04 | 19.6-6.4 |
| K.NH3 / µ.AMO / µ.HAO | | | | 0.10 - 0.0031 - 0.00080 | -0.5/0.4/0.67 | 25.2-5.3 |
| K.NH3 / µ.AMO / K.X | | | | 0.10 - 0.0031 - 0.16 | -0.88/0.92/-0.94 | 24.5-5.4 |
| K.NH3 / K.O2.AMO / µ.HAO | | | | 0.19 - 0.35 - 0.00033 | -0.86/-0.3/0.5 | 2.1-2.5 |
| K.NH3 / K.O2.AMO / K.X | | | | 0.20 - 3.6 - 0.10 | -0.26/0.8/-0.46 | 3.9-3.5 |
| K.NH3 / µ.HAO / K.X | | | | 0.18 - 0.00033 - 0.16 | 0.35/-0.4/-0.08 | 1.9-2.5 |
| K.O2.AMO / µ.AMO / µ.HAO | | | | 1.6 - 0.0060 - 0.00032 | -0.98/0.09/-0.16 | 2.6-3.6 |
| K.O2.AMO / µ.AMO / K.X | | | | 1.6 - 0.0050 - 0.10 | -0.47/0.35/-0.92 | 2.9-4.3 |
| K.O2.AMO / µ.HAO / K.X | | | | 0.29 - 0.00034 - 0.15 | 0.76/-0.89/-0.70 | 1.4-2.4 |
| µ.AMO / µ.HAO / K.X | | | | 0.0060 - 0.00032 - 0.16 | -0.78/0.25/-0.62 | 2.0-3.0 |

(*) Parameter subset selected for calibration. Maximum rates are shown as 1/minutes.

## Section S4 – Parameter estimation for aerobic $NH_2OH$ oxidation

### Modelling the HAO process in literature

In the past years $N_2O$ models describing $NH_4^+$ oxidation to $NO_2^-$ have included $NH_2OH$ as an obligate intermediate (Domingo-Félez and Smets, 2016; Ni et al., 2011; Ni et al., 2013a; Ni et al., 2014; Pocquet et al., 2016). Only recently the colorimetric method developed by (Frear and Burrell, 1955) has been applied to wastewater treating operations. $NH_2OH$ is not widely monitored and results from biological $NH_4^+$-removing systems have shown $NH_2OH$ levels between 0.01-0.15 mgN/L (Table S4). However, $N_2O$ models have not been calibrated with $NH_2OH$ datasets but with $NH_4^+$, $NO_2^-$, DO and less frequently NO.

**Table S4** – Reported $NH_2OH$ measurements from biological $NH_4^+$-removing systems.

| Reference | $NH_2OH$ (mgN/L) | System |
|---|---|---|
| (Soler-Jofra et al., 2016) | 0.03 - 0.11 | Full-scale SHARON (n = 5, Jan-Jun'15). Sampled during aeration. |
| (Udert et al., 2005) | < 0.14 | Lab-scale membrane bioreactor, influent NH4-NO2 (1:1). |
| (Harper et al., 2009) | ≈ 0* | Lab-scale nitrifying bioreactor, autotrophic biomass. |
| (De Clippeleir et al., 2013) | < 0.2 | OLAND, one-stage PNA. |
| (Kinh et al., 2016) | 0.01 - 0.11 - 0.15 | Lab-scale partial nitrification, SBR [pH 6.3 - 8.3 - 7.3] |
| (Yu and Chandran, 2010) | ≈ 0.03, max 0.05 | Batch culture exponential growth of Nitrosomonas sp. |

(*) data not reported

The hydroxylamine oxidoreductase (HAO) catalyses the $NH_2OH$ oxidation to $HNO_2$ with $H_2O$ as electron acceptor and not $O_2$ (Hooper and Terry, 1979; Sayavedra-Soto et al., 1996). The oxidation of $NH_2OH$ (specifically the further intermediate nitroxyl, HNO) to $HNO_2$ is assumed to involve $O_2$ because it does not occur under anoxia (Hooper and Terry, 1979). Still, the enzyme-catalysed aerobic conversion of NO to $HNO_2$ has not been demonstrated and the reduction of $O_2$ is believed to a parallel reaction occurring in cyt(aa3), a proton pumping terminal cytochrome oxidase (Figure S4-1) (Upadhyay et al., 2006). In recent $N_2O$ models the sequential $NH_2OH$ oxidation to $NO_2^-$ via NO is described with $O_2$ as electron acceptor and thus anoxic $NH_2OH$ oxidation does not occur (Ni et al., 2013a; Pocquet et al., 2016). Only the NDHA model can describe $N_2O$ production from both aerobic and anoxic $NH_2OH$ oxidation (with and without $NO_2^-$ production).

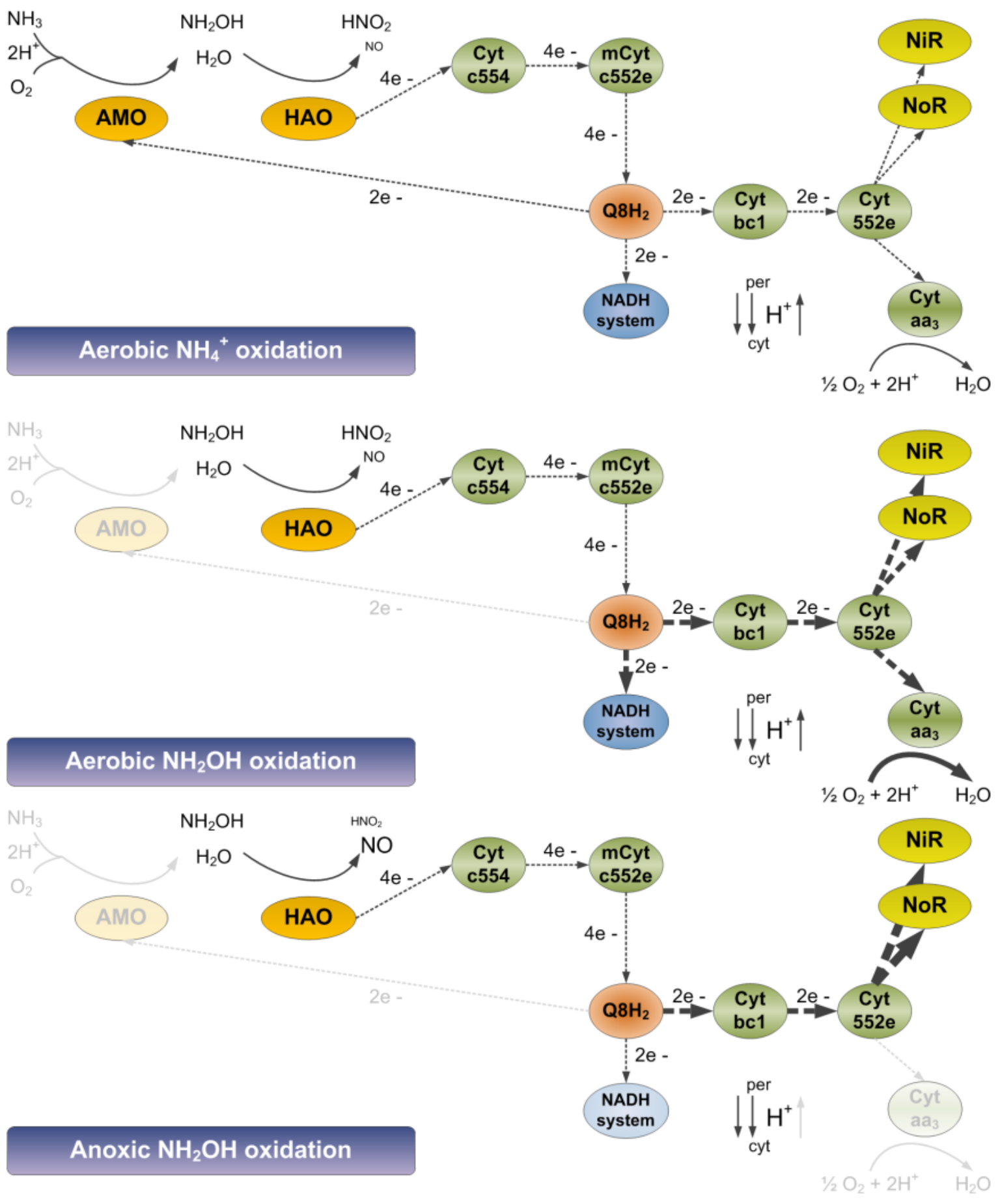

**Figure S4-1 -** Schematic of $NH_3$ oxidation to $HNO_2$ by AOB, main intermediates, electron flow and enzymatic sites.

Comparison of equivalent parameters from $N_2O$ models for the AMO and HAO process show high variability. $\mu_{MAX.HAO}$ has been reported as both higher and lower than $\mu_{AOB.AMO}$, ($\mu_{AOB.HAO}$ / $\mu_{AOB.AMO}$ = 0.3-2.6 (Ni et al., 2013a)). Substrate affinity $K_{NH2OH}$ has been assumed equal to that of $NH_4^+$ or has not been targeted during calibration (Law et al., 2011; Ni et al., 2011) with values ranging from 0.7 to 2.4 mgN/L (Ding et al., 2016; Law et al., 2011; Ni et al., 2011; Pocquet et al., 2016). The oxygen affinity of the HAO process ($K_{O2.HAO}$) has been reported at varying ratios compared to that of the AMO process ($K_{O2.HAO}/K_{O2.AMO}$ = 0.15 – 14) (Ni et al., 2013a; Ni et al., 2013b). Overall, while four new parameters were added to the model the information content of datasets remained invariant, which lowers parameter identifiability.

The AMO process cannot be faster than the HAO as the $NH_3$ oxidation process requires 2 electrons supplied by $NH_2OH$ oxidation, shown by an initial $NH_4^+$ oxidation acceleration phase (Chandran and Smets, 2008; Guisasola et al., 2006). Under steady-state, reported $K_{NH2OH}$ together with similar $\mu_{AOB.AMO}$, $\mu_{AOB.HAO}$, $K_{O2.AMO}$ and $K_{O2.HAO}$ yield $NH_2OH$ concentrations much larger than reported (≈ 1 mgN/L vs. 0.01-0.15mgN/L). The overestimation of $NH_2OH$ concentrations does not affect the overall steady state N-balance but biased $NH_2OH$ predictions directly affect $N_2O$ model calibrations because $NH_2OH$ is direct substrate of $N_2O$ production. The authors believe that even though describing $NH_2OH$ oxidation is still a challenge, $NH_2OH$ affinity should be higher than reported to predict more accurately the lower concentrations measured (Figure S4-2). $N_2O$ model calibrations will thus benefit from more accurate $NH_2OH$ predictions.

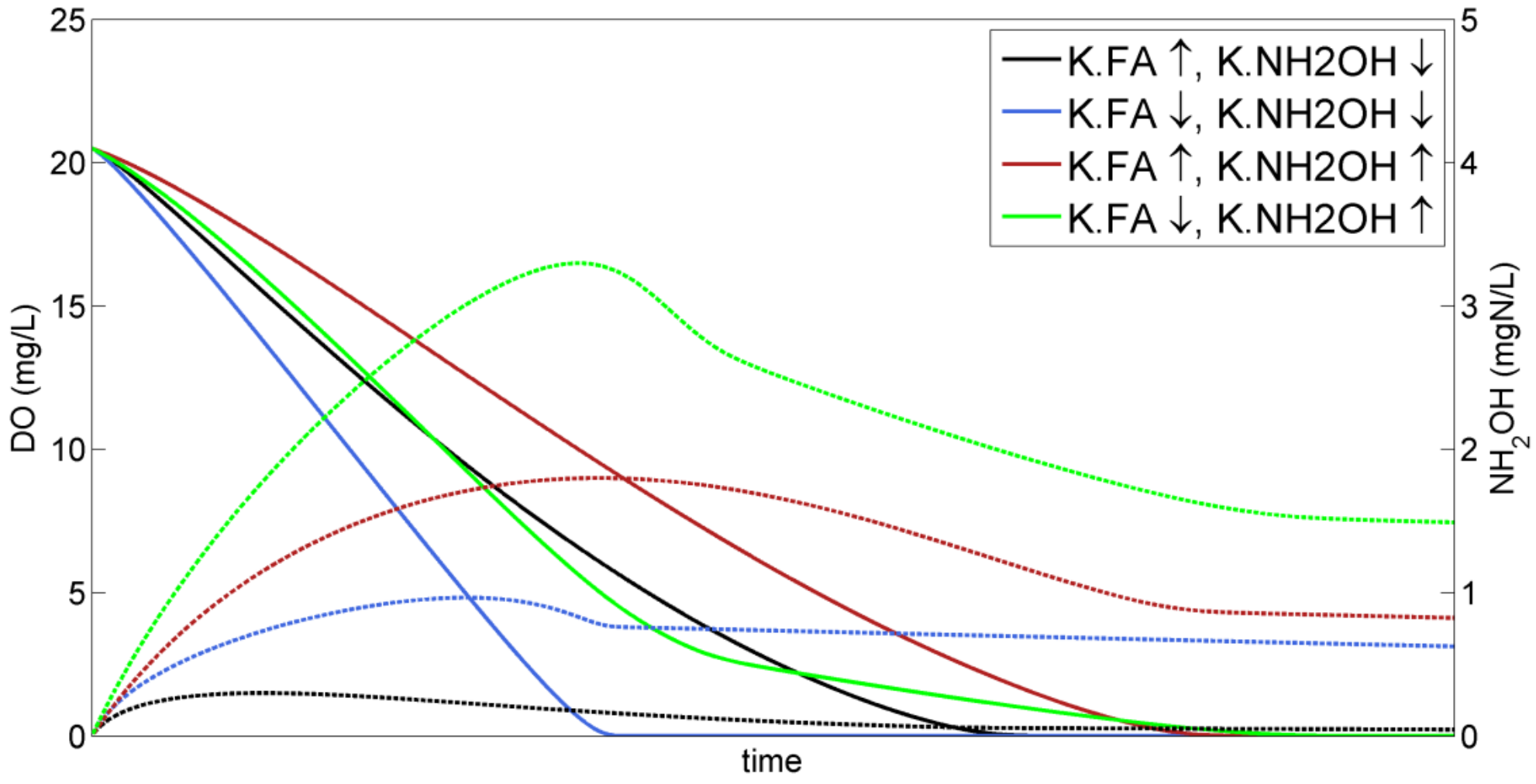

**Figure S4-2** – DO model simulation for an $NH_4^+$ pulse ($\mu_{MAX.AMO} = \mu_{MAX.HAO}$, $K_{O2.AMO} = K_{O2.HAO}$, pH = 7.5, 25 °C) to study the effect of $NH_3$ and $NH_2OH$ affinities to the DO and $NH_2OH$ concentration profiles. DO (straight lines), $NH_2OH$ (dotted line). ($K_{FA=NH3}$ = 0.01 – 0.1 mgN/L, $K_{NH2OH}$ = 0.2 – 2.4 mgN/L).

**Parameter estimation of Scen_HAO**

To determine $NH_2OH$ oxidation parameters independently from $NH_4^+$ oxidation (HAO vs. AMO) a similar approach to the 2-step nitrification procedure was used in this study (Brouwer et al., 1998). Based on the COD mass balance complete aerobic $NH_2OH$ removal was observed (Figure S4-3, t = 7 – 50min, 1.6mgN/L consumed, approximately 2.11 mgCOD/mgN). Moreover, from the independent pH dataset a decreasing trend during $NH_2OH$ removal agreed with the proposed reaction for the HAO process where $HNO_2$ is formed and mostly deprotonated at pH ≈ 7-8 (pKa = 3.26 at 25 °C). By adding $NH_2OH$ spikes at varying DO concentrations parameters describing $NH_2OH$ oxidation to $HNO_2$ in the NDHA model can be calibrated. From the sensitivity ranking the top-5 most sensitive parameters were considered for calibration: {$\mu_{AOB.HAO}$, $K_{AOB.NH2OH}$, $K_X$, $K_{AOB.NH3}$, $K_{HB.NH4}$} (Figure S4-3). Two parameters could be identified from this dataset: $\mu_{AOB.HAO}$, $K_{AOB.NH2OH}$ (Figure S4-4).

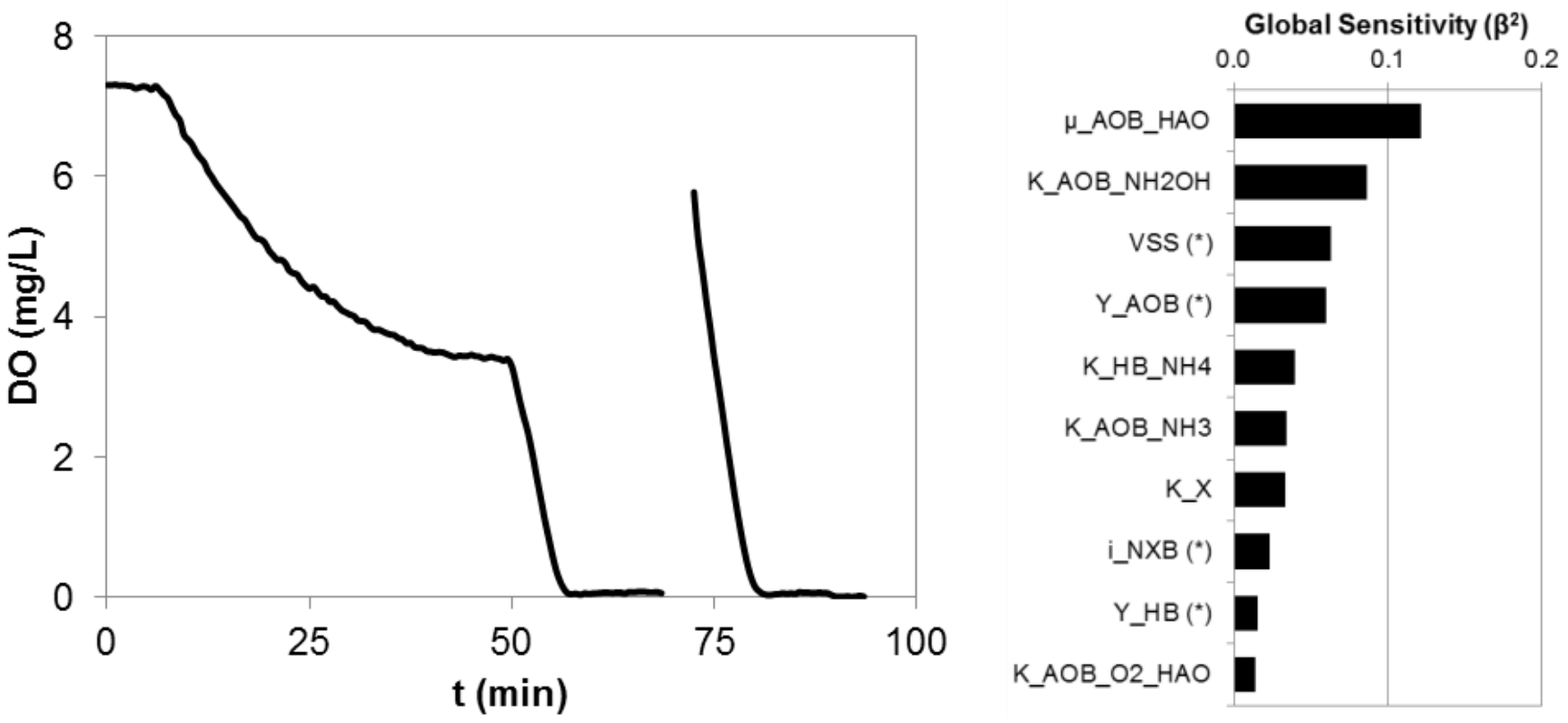

**Figure S4-3** - Left: DO measurements during $NH_2OH$ oxidation at varying $NH_2OH$ levels (1.6mgN/L t = 7min, 4mgN/L t = 50min, 4mgN/L t = 52min, 4mgN/L at t = 75 min). Right: Corresponding DO-associated sensitivity ranking.

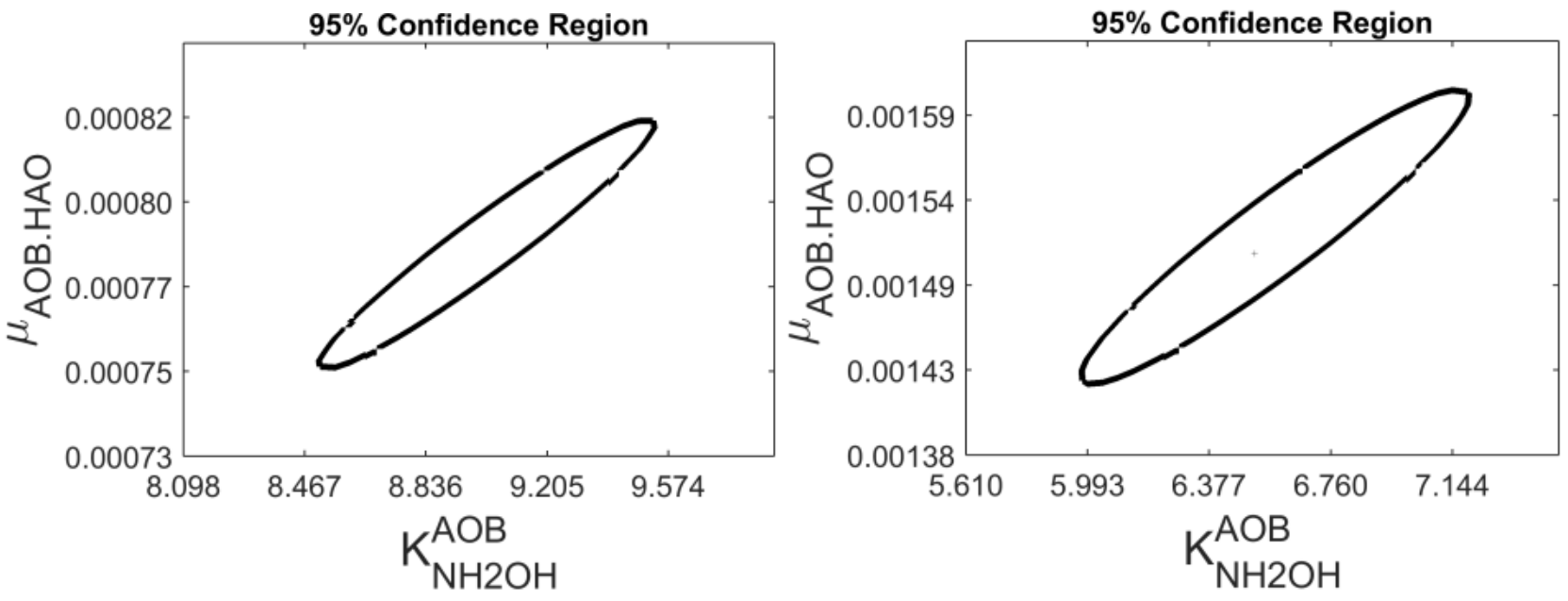

**Figure S4-4** - Calibration results from Scen_HAO. Left: $K_{AOB.NH2OH}$ and $\mu_{AOB.HAO}$ estimates for $Y_{AOB}$ = 0.18 ($K_{AOB.NH2OH}$ = 9.0mgN/L, $\mu_{AOB.HAO}$ = 0.00078 1/min, CV = 2.9-2.0%, corr = 0.93, RMSE = 0.07mg/L). Right: $K_{AOB.NH2OH}$ and $\mu_{AOB.HAO}$ estimates for $Y_{AOB}$ = 0.36 assuming all 4 electrons released used for AOB growth. ($K_{AOB.NH2OH}$ = 6.5mgN/L, $\mu_{AOB.HAO}$ = 0.0015 1/min, CV = 4.8-3.0%, corr = 0.90, RMSE = 0.11mg/L).

The model described the DO profile with a low RMSE = 0.07mg/L and F-test = 1, indicating a good model prediction. Even though the estimated parameters showed a high correlation (≈ 0.9) the confidence intervals were not wide (< 3%) and the collinearity index was lower than 10, indicating good identifiability ($\mu_{AOB.HAO}$ = 0.64 1/d, $K_{AOB.NH2OH}$ = 9.0 mgN/L). High correlation occurred due to an unexpected high value for $K_{AOB.NH2OH}$ compared to [$NH_2OH_{init}$], which leads to practical identifiability problems in the Monod kinetics (Holmberg, 1982). While $\mu_{AOB.HAO}$ is similar to other literature values, $K_{AOB.NH2OH}$ was higher than those reported in $N_2O$ models but in the same range as in a similar experimental design (4.9 mgN/L) (Chandran and Smets, 2008). For three different AOB species an approximately double growth yield on $NH_2OH$ compared to $NH_4^+$ was observed by (Böttcher and Koops, 1994), as twice the electrons are available for growth. Moreover, more ATP was produced by AOA when $NH_2OH$ was fed compared to $NH_4^+$.

This modification was also considered in the calibration process but did not change the low affinity for $NH_2OH$ ($Y_{AOB}$ = 0.18 and 0.36 mgCOD/mgN, Figure S4-4).

**Role of $NH_2OH$ as intermediate of $NH_4^+$ oxidation by AOB**

The estimated parameters for the HAO process during aerobic conditions ($\mu_{MAX..HAO}$, $K_{AOB.NH2OH}$) overestimates $NH_2OH$ concentrations during $NH_4^+$ spikes (> 1 mgN/L simulated vs. < 0.15 mgN/L in literature, Table S4). Thus, isolated aerobic $NH_2OH$ oxidation is not representative of $NH_2OH$ oxidation from $NH_4^+$ production as two of the four electrons released by $NH_2OH$ oxidation to $NO_2^-$ allow the $O_2$ reduction/$NH_4^+$ oxidation of the AMO process. When $NH_4^+$ is not present these two electrons are not required and are thus channelled towards the cytochrome pool, which explains a much higher $N_2O$ yield on $NH_2OH$ compared to $NH_4^+$ (Figure 2) (de Bruijn et al., 1995; Cantera and Stein, 2007).

Overall, these findings show that a lower $K_{AOB.NH2OH}$ is necessary to describe the low $NH_2OH$ accumulation found in literature where, under steady-state, the AMO and HAO processes occur at the same rate. If $NH_2OH$ accumulated after an $NH_4^+$ spike due to a slower HAO to AMO process the oxygen consumption rate would decrease from 3.43 to 2.29 mgCOD/mgN after $NH_4^+$ depletion. The same change in oxygen consumption occurs in NOB-limited nitrifying respirograms where $NO_2^-$ is still present after $NH_4^+$ has been depleted (Chandran and Smets, 2000). Moreover, the release in high concentrations of the main electron sourcewould be, potentially, a waste of energy for AOB.

## Section S5 – Parameter estimation results: Validation of model response and parameter estimates.

### Normality of error distribution

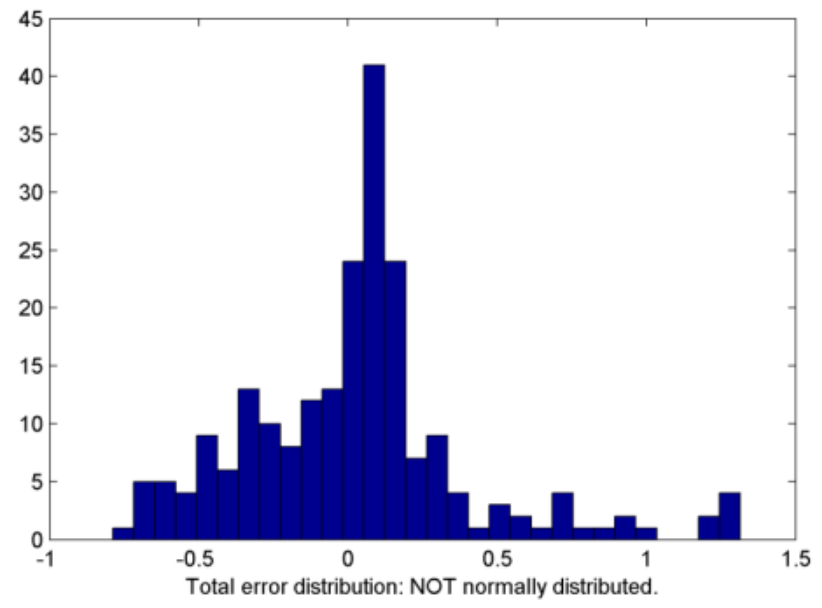

**Figure S5-1**. Distribution of residuals of Scen_AMO.

### Data acquisition frequency impact on calibration results:

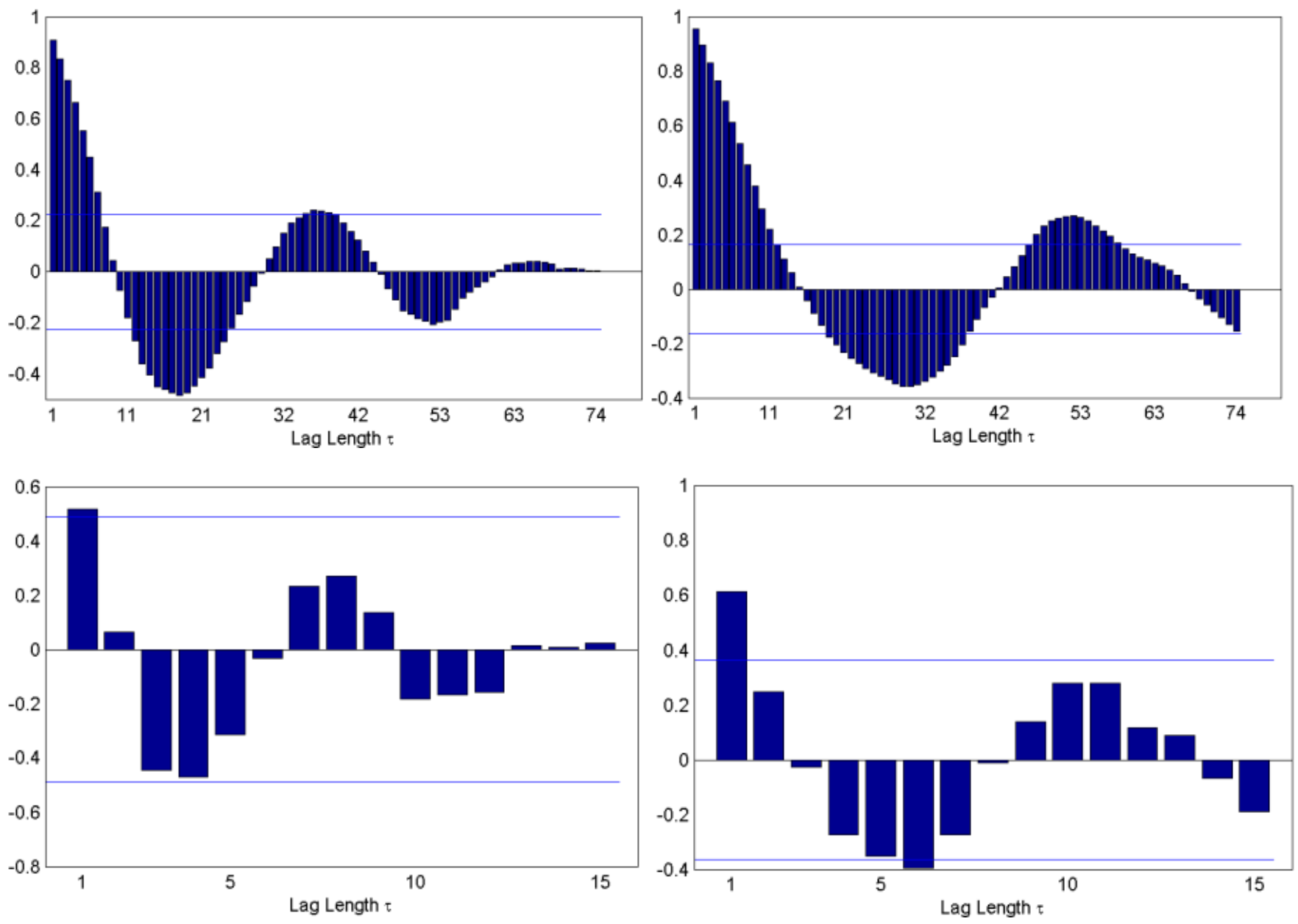

**Figure S5-2**. Autocorrelation function of residuals from best-fit estimates of two experiments from Scen_AMO. Data frequency = point/2min (top), point/10min (bottom).

| | J_crit | Error | $\mu_{NOB}$ (1/min) | CV $\mu_{NOB}$ (%) | $k_H$ | CV $k_H$ (%) | Corr | F-test | Nr. points |
|---|---|---|---|---|---|---|---|---|---|
| Sampling = 0.5 min | 1.01 | 5.95 | 0.000670 | 1.0 | 0.0017 | 0.9 | -0.52 | 1 (2.9/3) | 1020 |
| Sampling = 2 min | 1.04 | 5.88 | 0.000666 | 2.1 | 0.0017 | 1.8 | -0.51 | 1 (0.8/3) | 257 |
| Sampling = 5 min | 1.10 | 5.97 | 0.000669 | 3.4 | 0.0017 | 3.0 | -0.52 | 1 (0.3/3) | 104 |
| Sampling = 10 min | 1.20 | 5.74 | 0.000661 | 4.8 | 0.0017 | 4.2 | -0.51 | 1 ( 0.2/3) | 53 |

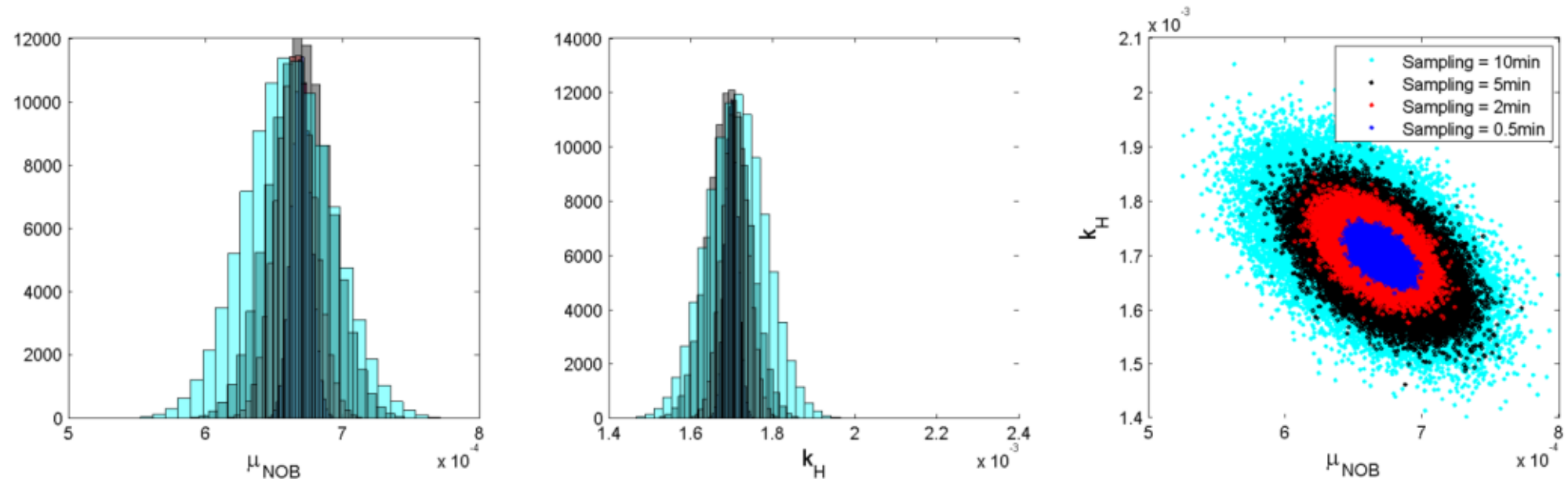

**Figure S5-3**. Samples from $\mu_{NOB}$ and $k_H$ estimates for different data frequencies used for calibration (every 0.5-2-5-10 minutes).

**Table S5-1** – Analysis of residuals for DO and $N_2O$ calibration and validation.

**DO calibration**

| Scenario | Scen_NOB | | | Scen_AMO | | | Scen_AMO_DO | | |
|---|---|---|---|---|---|---|---|---|---|
| Parameters Estimated | $\mu_{NOB}$, $k_H$ | | | $\mu_{AOB.AMO}$, $K_{AOB.NH3}$ | | | $K_{AOB.O2.AMO}$ | | |
| | Exp_1 | Exp_2 | Combined | Exp_1 | Exp_2 | Combined | Exp_1 | Exp_2 | Combined |
| R2 | 0.9986 | 0.9964 | 0.9958 | 0.9989 | 0.9987 | 0.9987 | 0.9983 | 0.9982 | 0.9982 |
| Number_points | 86 | 171 | 257 | 75 | 142 | 217 | 34 | 61 | 95 |
| MSEP | 0.0734 | 0.1618 | | 0.0673 | 0.1909 | | 0.0048 | 0.0079 | |
| ME | 0.4179 | 0.1169 | | 0.0072 | 0.0138 | | 0.0561 | 0.0317 | |
| SE | 0.1490 | 0.0707 | | 0.0136 | 0.0236 | | 0.0317 | 0.0014 | |
| NC | 0.4388 | 0.8175 | | 0.9826 | 0.9696 | | 0.9408 | 0.9831 | |
| RMSE | | | 0.3651 | | | 0.3868 | | | 0.0834 |
| F-test | | | 1 | | | 1 | | | 1 |

**DO validation**

| Scenario | Scen_AMO | | | | |
|---|---|---|---|---|---|
| | Exp_1 | Exp_2 | Exp_3 | Exp_4 | Exp_5 |
| R2 | 0.9981 | 0.9982 | 0.9947 | 0.9979 | 0.9964 |
| Number_points | 396 | 61 | 214 | 185 | 86 |
| MSEP | 0.2217 | 0.0078 | 0.6236 | 0.1697 | 0.0227 |
| ME | 0.1021 | 0.0274 | 0.2778 | 0.0425 | 0.1888 |
| SE | 0.2531 | 0.0014 | 0.2132 | 0.0236 | 0.4579 |
| NC | 0.6447 | 0.9874 | 0.5124 | 0.9391 | *0.3628* |
| RMSE | 0.4721 | 0.0897 | 0.7934 | 0.4142 | 0.1524 |
| F-test | *0* | 1 | *0* | 1 | *0* |

**$N_2O$ calibration**

| Scenario | Scen_An_HB | | | Scen_AMO | | | | Scen_AMO, Scen_An_AOB | | | |
|---|---|---|---|---|---|---|---|---|---|---|---|
| Parameters Estimated | $\eta_{HD}$ | | | $\varepsilon_{AOB}$ | | | | $\eta_{NOR}$, $K_{AOB.HNO2}$, $K_{AOB.NH2OH.ND}$ | | | |
| | Exp_1 | Exp_2 | Combined | Exp_1 | Exp_2 | Exp_3 | Combined | Exp_1 | Exp_2 | Exp_3 | Combined |
| R2 | 0.9783 | 0.9869 | 0.9779 | 0.6503 | 0.6211 | 0.9255 | 0.9974 | 0.9892 | 0.9728 | 0.998 | 0.9979 |
| Number_points | 208 | 240 | 448 | 315 | 230 | 192 | 737 | 284 | 214 | 70 | 568 |
| MSEP | 0.0022 | 0.0003 | | 0.0001 | 0.0001 | 0.0001 | | 0.0001 | 0.0001 | 0.0001 | |
| ME | 0.0664 | 0.3456 | | 0.1331 | 0.1515 | 0.6136 | | 0.1461 | 0.0073 | 0.0300 | |
| SE | 0.0030 | 0.0767 | | 0.0622 | 0.3636 | 0.0002 | | 0.3661 | 0.7190 | 0.0266 | |
| NC | 0.9350 | 0.5805 | | 0.8074 | 0.4886 | *0.3882* | | 0.4909 | *0.2784* | 0.9574 | |
| RMSE | | | 0.0342 | | | | 0.0007 | | | | 0.0019 |
| F-test | | | 1 | | | | 1 | | | | 1 |

**$N_2O$ validation**

| Scenario | Scen_AMO, Scen_An_AOB | | |
|---|---|---|---|
| | Exp_1 | Exp_2 | Exp_3 |
| R2 | 0.9618 | 0.9692 | 0.9958 |
| Number_points | 396 | 261 | 60 |
| MSEP | 0.0001 | 0.1416 | 0.0001 |
| ME | 0.3234 | 0.4641 | 0.2421 |
| SE | 0.0526 | 0.0182 | 0.4015 |
| NC | 0.625 | 0.5198 | 0.3692 |
| RMSE | 0.0011 | 0.3777 | 0.0033 |
| F-test | 1 | *0* | 1 |

## Section S6 – AOB-driven $N_2O$ production from primary N-substrates

### Biotically driven $N_2O$ production associated to $NH_4^+$ oxidation

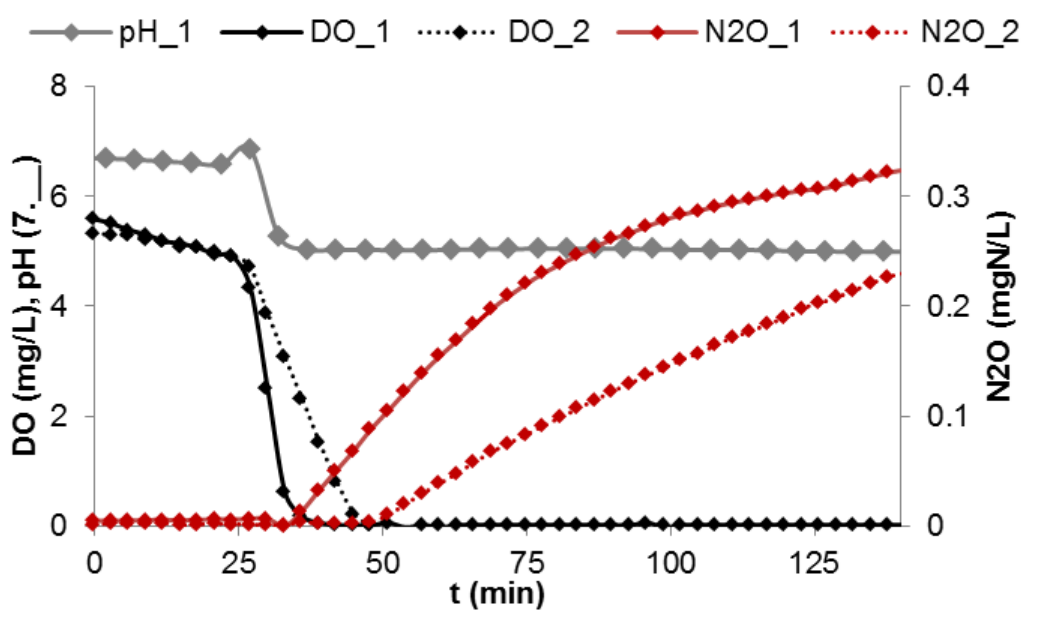

**Figure S6-1 –** Biotically-driven $N_2O$ production from $NH_4^+$ oxidation. VSS_1 = 0.3, VSS_2 = 0.13g/L yielded similar specific $N_2O$ production rates: N$_2$O_R1 = 0.21 and N$_2$O_R2 = 0.24 mgN2O-N/gVSS.min.

### Role of $NH_4^+$, $NO_2^-$ and $NH_2OH$ on anoxic $N_2O$ production

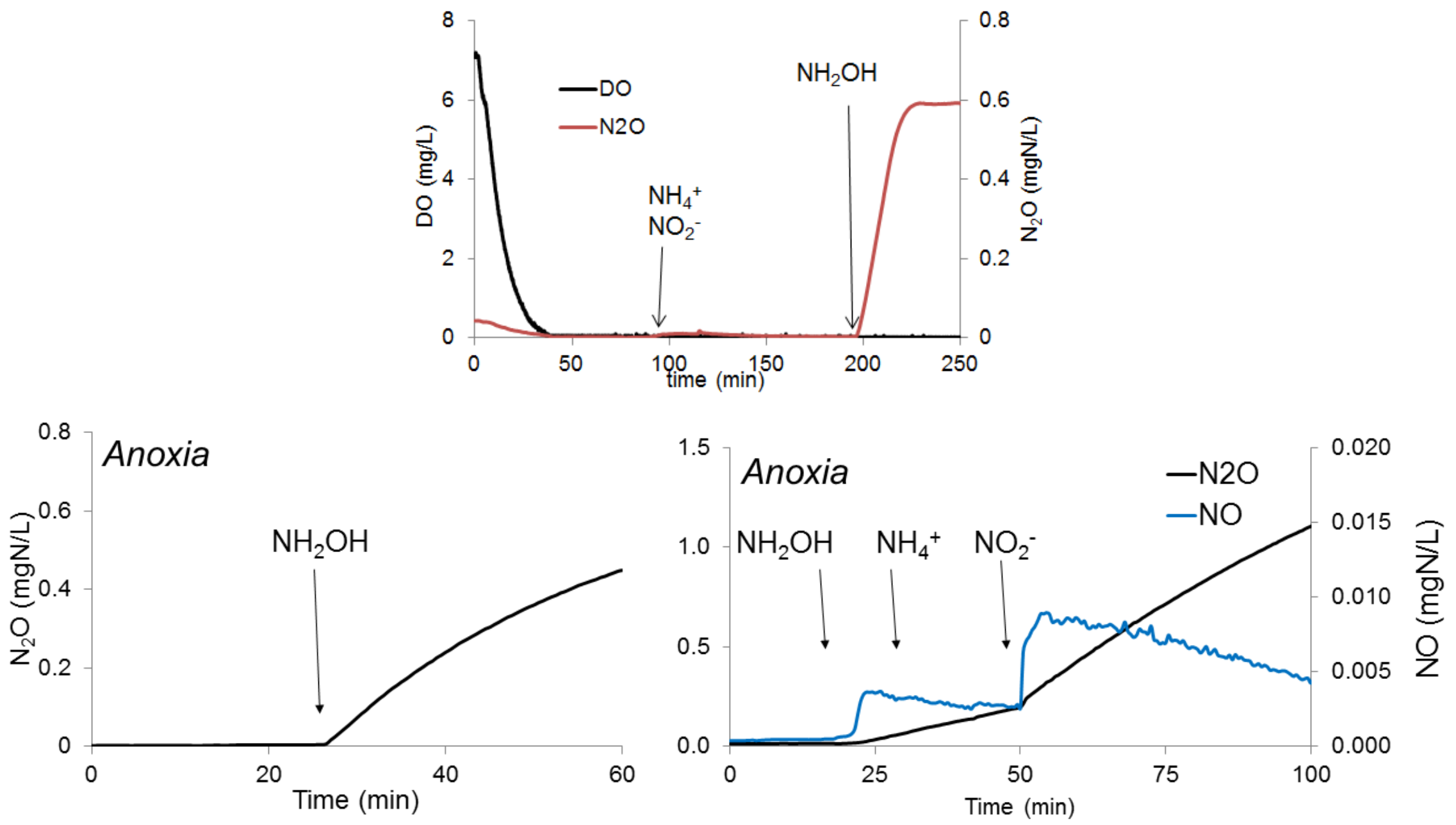

**Figure S6-2 –** (Top) Biotically-driven $N_2O$ production from anoxic $NH_2OH$ oxidation. (Bottom) $N_2O$ and NO liquid concentrations from anoxic $NH_2OH$ oxidation (left, $NO_2^-$ < 0.05 mgN/L at the end of the experiment), and anoxic $NH_2OH$ oxidation followed by $NH_4^+$ and $NO_2^-$ addition (right).

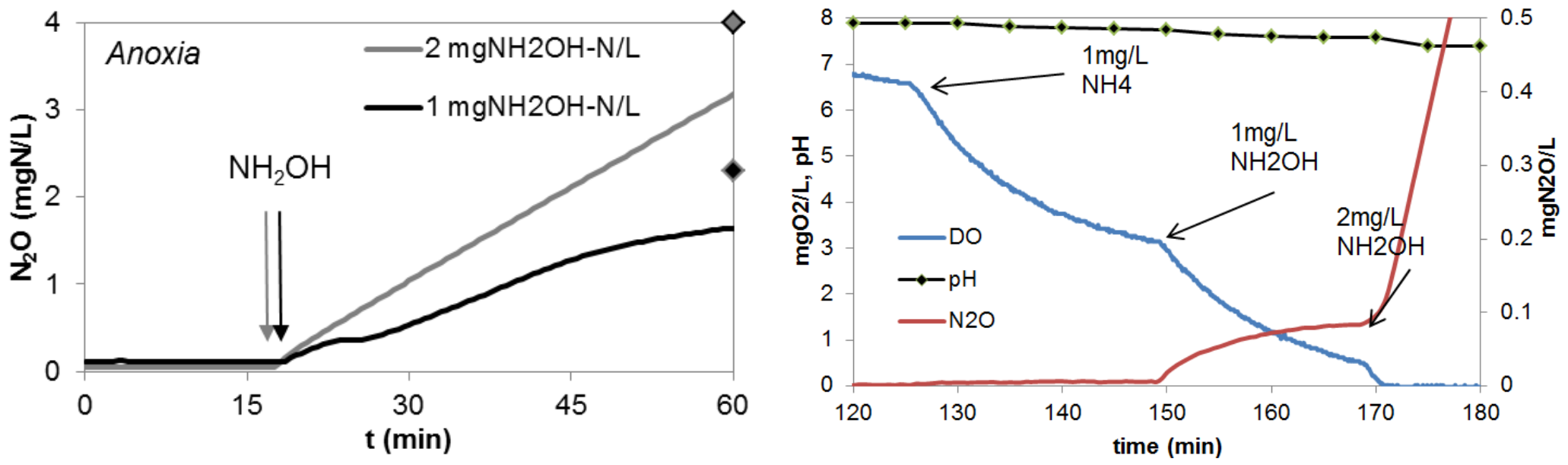

**Figure S6-3 –** (Left) $N_2O$ concentration after anoxic $NH_2OH$ spikes under excess $NO_2^-$ (Markers, final $N_2O$ concentrations). (Rigth) DO, pH and $N_2O$ profiles after aerobic $NH_4^+$ (1mgN/L), $NH_2OH$ (1 and 2 mgN/L) spikes (Scen_HAO).

Under $NO_2^-$ excess a $NH_2OH$ spike yielded more $N_2O$ than the corresponding $NH_2OH$ added, indicating a combined $NH_2OH$ and $NO_2^-$ anoxic reaction (**Figure S6-4**).

**Table S6-1 –** $N_2O$ production rates from two duplicate experiments of aerobic $NH_4^+$ pulses, aerobic $NH_2OH$ pulses, low DO $NH_2OH$ pulses, anoxic $NH_2OH$ pulses.

**N2O_R (mgN2O-N/gVSS.min)**

| Pulse | NG1 | NG2 |
|---|---|---|
| 1 NH4 | 0.0008 | 0.0007 |
| 1 NH2OH | 0.027 | 0.033 |
| 2 NH2OH _@_limit_DO | 0.023 | 0.033 |
| < 2 NH2OH_anox | 0.22 | 0.13 |

**Abiotic $N_2O$ production**

To study the effect of $HNO_2$, $NH_2OH$ and pH on abiotic $N_2O$ production a factorial experimental design is constructed (Table S6-2). Results showed that in the absence of $NO_2^-$, $NH_2OH$-driven abiotic $N_2O$ production only occurs at very high pH (≥ 8.7) (Figure S6-4). Coupling $HNO_2$ and $NH_2OH$ produced $N_2O$ at: high pH (≥ 8) + high $NH_2OH$ (≥ 0.5 mgN/L). Therefore high $NO_2^-$ and $NH_2OH$ concentrations are necessary, outside the range of typical wastewater systems (pH > 8.4, $NO_2^-$ > 500 mgN/L, $NH_2OH$ ≥ 0.5 mgN/L).

**Table S6-2 -** Factorial experimental design to study abiotic $N_2O$ production.

| | | | | | |
|---|---|---|---|---|---|
| $HNO_2$ (µgN/L) | 0 | 0.2 | 2 | 20 | 100 |
| $NH_2OH$ (mgN/L) | 0 | 0.05 | 0.2 | 0.5 | 2 |
| pH | 6.5 | 7.25 | 8 | 8.7 | 9.4 |

Overall, the substrate concentrations necessary to produce $N_2O$ abiotically are outside the range of the experiments design to calibrate the NDHA model: high pH, high $NO_2^-$ and high $NH_2OH$.

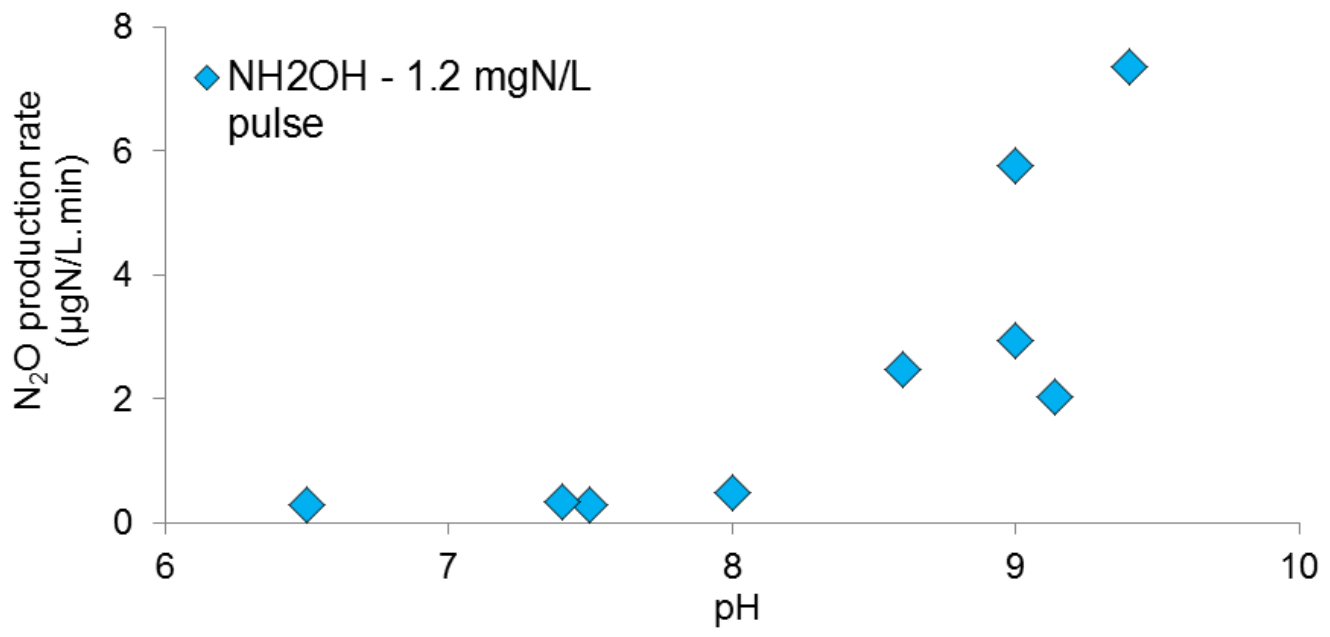

**Figure S6-4 -** Abiotic $N_2O$ production rates for $NH_2OH$ pulses (1.2 mgN/L) at varying pH.

## Section S7 – Pathway contribution to total $N_2O$ pool during model fitting

Based on the calibrated NDHA model the contribution of each pathway to the total $N_2O$ production was analysed under various scenarios. During $NH_4^+$ oxidation at high DO the NN pathway was responsible for all the $N_2O$ produced as the ND and ND pathways are highly inhibited (Figure S7-1). The NN contribution is directly related to $NH_2OH$, thus when $NH_4^+$ is depleted the NN contribution tapers off. As the DO level decreased the contribution of ND increased as both $NH_2OH$ and FNA were present. All the $NH_2OH$ left after reaching anoxia yielded $N_2O$ from both the NN and ND pathways. Even under anoxia the NN pathway is still active but at a much lower rate than the ND pathway which clearly outcompeted the $NH_2OH$ uptake in the presence of FNA, becoming the dominant $N_2O$ producing pathway in the transition to anoxia. The HD pathway followed a similar trend to ND but at much lower rates as no external carbon was added.

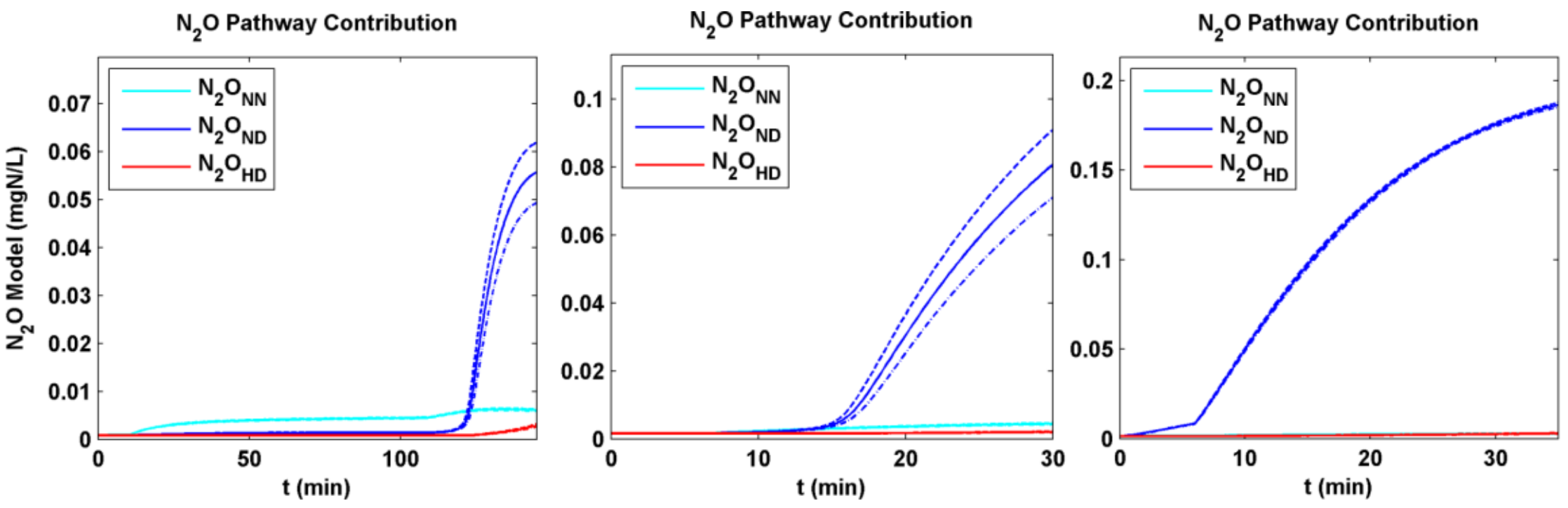

**Figure S7-1 –** Pathway contribution to the total $N_2O$ pool and 95%CI from calibrated parameters for: (Left, Middle) Scen_AMO ($NH_4^+$ pulses, Aerobic → anoxic), (Right) Scen_HAO ($NH_2OH$ pulse, anoxic).

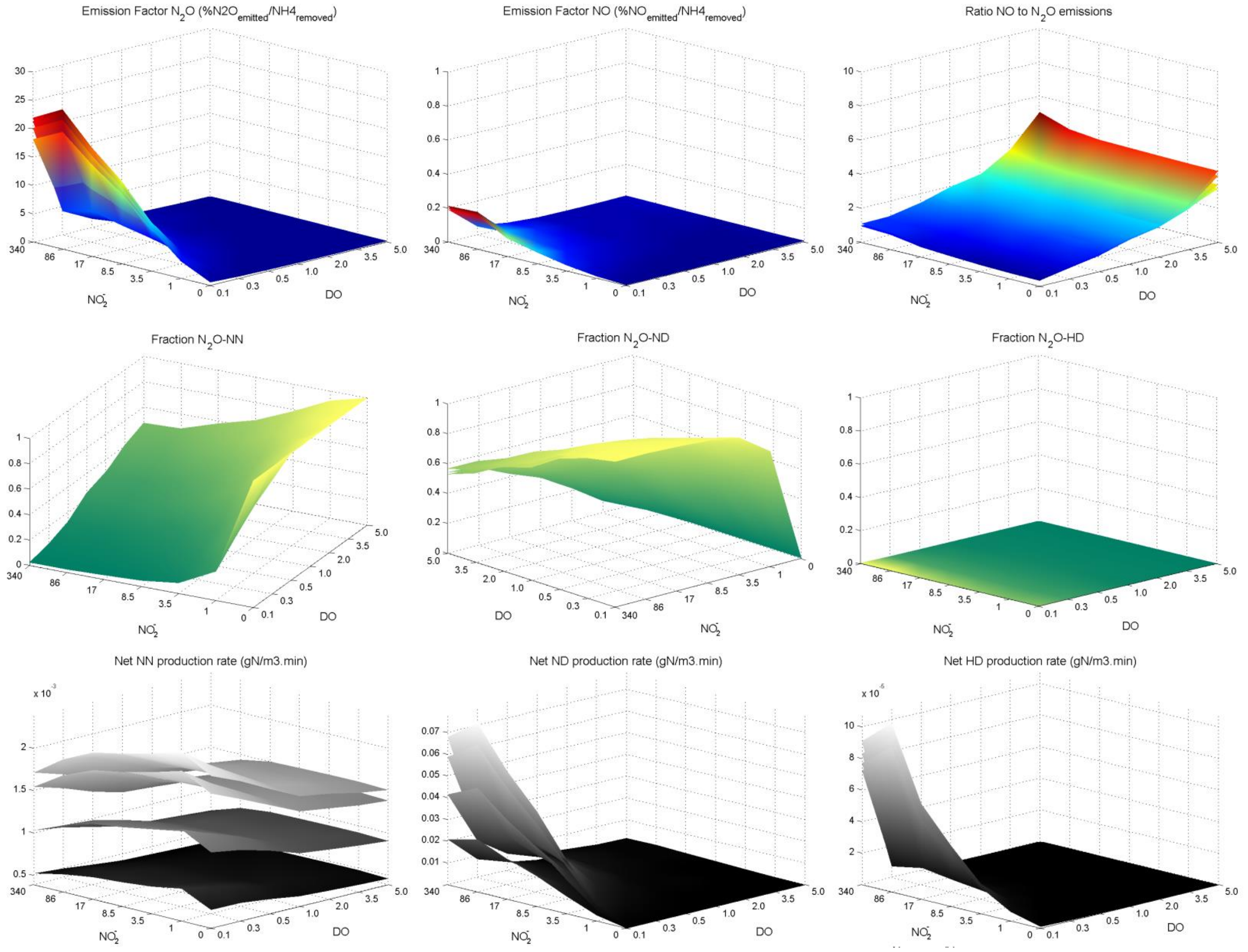

## Section S8 – NDHA model evaluation

**Figure S8-1 –** Scenario analysis for model evaluation with best-fit parameters. Simulations were run under constant DO (0.1 - 0.3 - 0.5 - 1.0 - 2.0 - 3.5 - 5.0 mg/L), $NO_2^-$ (0 – 1 – 3.5 – 8.5 – 17 – 86 – 340 mgN/L) and $NH_4^+$ (2.4 – 7.0 – 24 – 70 mgN/L). Top: (from left to right) $N_2O$, NO and NO/$N_2O$ emission factor (% $N_2O/NH_4^+$). Middle: NN, ND and HD Pathway contribution ( - ). Bottom: Net NN, ND and HD production rate ($gN/m^3$.min).

## Section S9 – Uncertainty propagation during $NH_4^+$ oxidation

The default uncertainty was propagated for all model parameters and compared to the uncertainty of only the calibrated parameters ($\mu_{NOB}$, $k_H$, $\mu_{AOB.AMO}$, $K_{AOB.NH3}$ and $K_{AOB.O2.AMO}$). A similar width of the 95% CI indicate that the model outputs were very sensitive to the parameter subset calibrated. After calibration the uncertainty obtained for those five parameters was also propagated highlighting a significant decrease of the prediction uncertainty, not only for the main N-species but also for $N_2O$. A precise calibration of $NH_4^+$ removal will help reduce the uncertainty associated to $N_2O$ production, which will benefit the experimental designs focused on $N_2O$ calibrations.

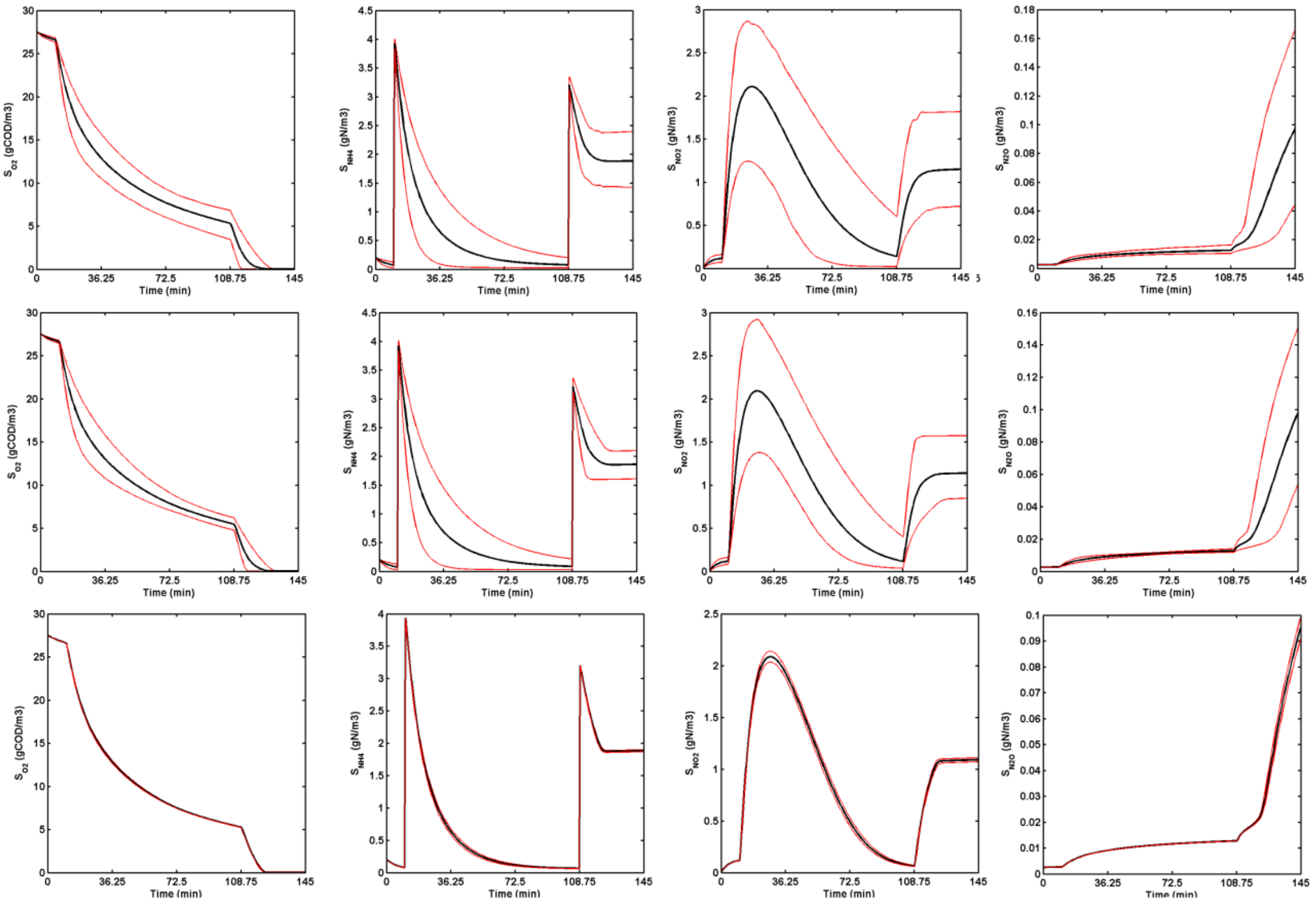

**Figure S9-1 –** Uncertainty propagation for an experiment of Scen_AMO (x2 $NH_4^+$ pulses) Top panels: Default uncertainty for all model parameters. Middle: Default uncertainty only for parameters estimated with the DO dataset ($\mu_{NOB}$, $k_H$, $\mu_{AOB.AMO}$, $K_{AOB.NH3}$ and $K_{AOB.O2.AMO}$). Bottom: Uncertainty from DO calibration only for parameters estimated with the DO dataset (Table 2).

## Section S10 – NDHA model

| Component (i) ▸ / Process (j) ▾ | | 1 $S_S$ | 2 $S_{O2}$ | 3 $S_{NH3}$ | 4 $S_{NH2OH}$ | 5 $S_{HNO2}$ | 6 $S_{NO3}$ | 7 $S_{NO}$ | 8 $S_{N2O}$ | 9 $S_{N2}$ | 10 $S_{IC}$ | 11 $X_{B,AOB}$ | 12 $X_{B,NOB}$ | 13 $X_{B,H}$ | 14 $X_S$ | 15 $X_I$ |
|---|---|---|---|---|---|---|---|---|---|---|---|---|---|---|---|---|
| ***AOB growth*** | | | | | | | | | | | | | | | | |
| Aerobic_AMO | 1 | | -1.14 | -1 | 1 | | | | | | -1/14 | | | | | |
| Aerobic_HAO* | 2 | | | $-i_{NXB}$ | $-\frac{1}{Y_{AOB}}$ | | | $\frac{1}{Y_{AOB}}$ | | | $-1/14 \cdot (i_{NXB})$ | 1 | | | | |
| Aerobic_HAO | 3 | | $-\left(\frac{2.29-Y_{AOB}}{Y_{AOB}}\right)$ | $-i_{NXB}$ | $-\frac{1}{Y_{AOB}}$ | $\frac{1}{Y_{AOB}}$ | | | | | $-1/14 \cdot (i_{NXB}-1/Y_{AOB})$ | 1 | | | | |
| Anox_A_NIR | 4 | | | | -1 | -3 | | 4 | | | 3/14 | | | | | |
| Anox_A_NOR | 5 | | | | -1 | | | -2 | 3 | | -1/14 | | | | | |
| ***NOB growth*** | | | | | | | | | | | | | | | | |
| Aer_NOB_growth | 6 | | $-\left(\frac{1.14-Y_{NOB}}{Y_{NOB}}\right)$ | $-i_{NXB}$ | | $-\frac{1}{Y_{NOB}}$ | $\frac{1}{Y_{NOB}}$ | | | | $-i_{NXB}/14$ | | 1 | | | |
| ***HB growth*** | | | | | | | | | | | | | | | | |
| Aerobic_H_growth | 7 | $-\frac{1}{Y_{HB}}$ | $-\left(\frac{1-Y_{HB}}{Y_{HB}}\right)$ | $-i_{NXB}$ | | | | | | | $-i_{NXB}/14$ | | | 1 | | |
| Anox_H_NAR | 8 | $-\frac{1}{Y_{HB}}$ | | $-i_{NXB}$ | | $\left(\frac{1-Y_{HB}}{1.14 \cdot Y_{HB}}\right)$ | $-\left(\frac{1-Y_{HB}}{1.14 \cdot Y_{HB}}\right)$ | | | | $-i_{NXB}/14$ | | | 1 | | |
| Anox_H_NIR | 9 | $-\frac{1}{Y_{HB}}$ | | $-i_{NXB}$ | | $-\left(\frac{1-Y_{HB}}{0.57 \cdot Y_{HB}}\right)$ | | $\left(\frac{1-Y_{HB}}{0.57 \cdot Y_{HB}}\right)$ | | | $-1/14 \cdot (i_{NXB} - (1-Y_{HB}) / (0.57 \cdot Y_{HB})$ | | | 1 | | |
| Anox_H_NOR | 10 | $-\frac{1}{Y_{HB}}$ | | $-i_{NXB}$ | | | | $-\left(\frac{1-Y_{HB}}{0.57 \cdot Y_{HB}}\right)$ | $\left(\frac{1-Y_{HB}}{0.57 \cdot Y_{HB}}\right)$ | | $-i_{NXB}/14$ | | | 1 | | |
| Anox_H_NOS | 11 | $-\frac{1}{Y_{HB}}$ | | $-i_{NXB}$ | | | | | $-\left(\frac{1-Y_{HB}}{0.57 \cdot Y_{HB}}\right)$ | $\left(\frac{1-Y_{HB}}{0.57 \cdot Y_{HB}}\right)$ | $-i_{NXB}/14$ | | | 1 | | |
| ***Lysis*** | | | | | | | | | | | | | | | | |
| AOB | 12 | | | $i_{NXB}-f_I.i_{NXI}-(1-f_I).i_{NXS}$ | | | | | | | $(i_{NXB}-f_I.i_{NXI}-(1-f_I).i_{NXS})/14$ | -1 | | | $1-f_{XI}$ | $f_{XI}$ |
| NOB | 13 | | | $i_{NXB}-f_I.i_{NXI}-(1-f_I).i_{NXS}$ | | | | | | | $(i_{NXB}-f_I.i_{NXI}-(1-f_I).i_{NXS})/14$ | | -1 | | $1-f_{XI}$ | $f_{XI}$ |
| HB | 14 | | | $i_{NXB}-f_I.i_{NXI}-(1-f_I).i_{NXS}$ | | | | | | | $(i_{NXB}-f_I.i_{NXI}-(1-f_I).i_{NXS})/14$ | | | -1 | $1-f_{XI}$ | $f_{XI}$ |
| ***Hydrolysis*** | | | | | | | | | | | | | | | | |
| Aerobic | 15 | 1 | | $i_{NXS}$ | | | | | | | $i_{NXS}/14$ | | | | -1 | |
| Anoxic | 16 | 1 | | $i_{NXS}$ | | | | | | | $i_{NXS}/14$ | | | | -1 | |
| Anaerobic | 17 | 1 | | $i_{NXS}$ | | | | | | | $i_{NXS}/14$ | | | | -1 | |
| *Aeration* | 18 | | 1 | | | | | | | | | | | | | |
| *Stripping_N2O* | 19 | | | | | | | | -1 | | | | | | | |
| *Stripping_NO* | 20 | | | | | | | -1 | | | | | | | | |

| Component (i) ▸ / *Process (j)* ▼ | | Process Rate (g·m$^{-3}$·s$^{-1}$) |
|---|---|---|
| ***AOB growth*** | | |
| Aerobic_AMO | *1* | $\mu_{AMO}^{AOB}\cdot\frac{S_{O2}}{S_{O2}+K_{O2_AMO}^{AOB}}\cdot\frac{S_{NH3}}{S_{NH3}+K_{NH3}^{AOB}+S_{NH3}^{2}/K_{i_NH3}^{AOB}}\cdot\frac{K_{i_HNO2}^{AOB}}{S_{HNO2}+K_{i_HNO2}^{AOB}}\cdot X_{AOB}$ |
| Aerobic_HAO* | *2* | $\mu_{HAO}^{AOB}\cdot\varepsilon_{AOB}\cdot\frac{S_{NH2OH}}{S_{NH2OH}+K_{NH2OH}^{AOB}}\cdot X_{AOB}$ |
| Aerobic_HAO | *3* | $\mu_{HAO}^{AOB}\cdot(1-\varepsilon_{AOB})\cdot\frac{S_{O2}}{S_{O2}+K_{O2_HAO}^{AOB}}\cdot\frac{S_{NH2OH}}{S_{NH2OH}+K_{NH2OH}^{AOB}}\cdot X_{AOB}$ |
| Anox_A_NIR | *4* | $\mu_{HAO}^{AOB}\cdot\eta_{NIR}\cdot\frac{K_{i_O2}^{AOB}}{S_{O2}+K_{i_O2}^{AOB}}\cdot\frac{S_{NH2OH}}{S_{NH2OH}+K_{NH2OH_ND}^{AOB}}\cdot\frac{S_{HNO2}}{S_{HNO2}+K_{HNO2}^{AOB}}\cdot X_{AOB}$ |
| Anox_A_NOR | *5* | $\mu_{HAO}^{AOB}\cdot\eta_{NOR}\cdot\frac{S_{NH2OH}}{S_{NH2OH}+K_{NH2OH_ND}^{AOB}}\cdot\frac{S_{NO}}{S_{NO}+K_{NO}^{AOB}}\cdot X_{AOB}$ |
| ***NOB growth*** | | |
| Aer_NOB_growth | 6 | $\mu_{NOB}\cdot\frac{S_{O2}}{S_{O2}+K_{O2}^{NOB}}\cdot\frac{S_{HNO2}}{S_{HNO2}+K_{HNO2}^{NOB}+S_{HNO2}^{2}/K_{i_HNO2}^{NOB}}\cdot\frac{K_{i_NH3}^{NOB}}{S_{NH3}+K_{i_NH3}^{NOB}}\cdot X_{NOB}$ |
| ***HB growth*** | | |
| Aerobic_H_growth | 7 | $\mu_{HB}\cdot\frac{S_{O2}}{S_{O2}+K_{O2}^{HB}}\cdot\frac{S_{NH4}}{S_{NH4}+K_{NH4}^{HB}}\cdot\frac{S_{S}}{S_{S}+K_{S}^{HB}}\cdot X_{HB}$ |
| Anox_H_NAR | 8 | $\mu_{NAR}^{HB}\cdot\frac{K_{i_O2_NAR}^{HB}}{S_{O2}+K_{i_O2_NAR}^{HB}}\cdot\frac{S_{S}}{S_{S}+K_{S_NAR}^{HB}}\cdot\frac{S_{NH4}}{S_{NH4}+K_{NH4}^{HB}}\cdot\frac{S_{NO3}}{S_{NO3}+K_{NO3}^{HB}}\cdot X_{HB}$ |
| Anox_H_NIR | 9 | $\mu_{NIR}^{HB}\cdot\frac{K_{i_O2_NIR}^{HB}}{S_{O2}+K_{i_O2_NIR}^{HB}}\cdot\frac{K_{i_NO_NIR}^{HB}}{S_{NO}+K_{i_NO_NIR}^{HB}}\cdot\frac{S_{S}}{S_{S}+K_{S_NIR}^{HB}}\cdot\frac{S_{NH4}}{S_{NH4}+K_{NH4}^{HB}}\cdot\frac{S_{NO2}}{S_{NO2}+K_{NO2}^{HB}}\cdot X_{HB}$ |
| Anox_H_NOR | 10 | $\mu_{NOR}^{HB}\cdot\frac{K_{i_O2_NOR}^{HB}}{S_{O2}+K_{i_O2_NOR}^{HB}}\cdot\frac{S_{S}}{S_{S}+K_{S_NOR}^{HB}}\cdot\frac{S_{NH4}}{S_{NH4}+K_{NH4}^{HB}}\cdot\frac{S_{NO}}{S_{NO}+K_{NO}^{HB}+S_{NO}{}^{2}/K_{i_NO_NOR}^{HB}}\cdot X_{HB}$ |
| Anox_H_NOS | 11 | $\mu_{NOS}^{HB}\cdot\frac{K_{i_O2_NOS}^{HB}}{S_{O2}+K_{i_O2_NOS}^{HB}}\cdot\frac{K_{i_NO_NOS}^{HB}}{S_{NO}+K_{i_NO_NOS}^{HB}}\cdot\frac{S_{S}}{S_{S}+K_{S_NOS}^{HB}}\cdot\frac{S_{NH4}}{S_{NH4}+K_{NH4}^{HB}}\cdot\frac{S_{N2O}}{S_{N2O}+K_{N2O}^{HB}}\cdot X_{HB}$ |
| ***Lysis*** | | |
| AOB | 12 | $b_{AOB}\cdot\left(\frac{S_{O2}}{S_{O2}+K_{O2_b}}+\eta_{b}\cdot\frac{K_{O2_b}}{K_{O2_b}+S_{O2}}\cdot\frac{S_{NOx}}{K_{NOx}+S_{NOx}}\right)\cdot X_{AOB}$ |
| NOB | 13 | $b_{NOB}\cdot\left(\frac{S_{O2}}{S_{O2}+K_{O2_b}}+\eta_{b}\cdot\frac{K_{O2_b}}{K_{O2_b}+S_{O2}}\cdot\frac{S_{NOx}}{K_{NOx}+S_{NOx}}\right)\cdot X_{NOB}$ |
| HB | 14 | $b_{HB}\cdot\left(\frac{S_{O2}}{S_{O2}+K_{O2_b}}+\eta_{b}\cdot\frac{K_{O2_b}}{K_{O2_b}+S_{O2}}\cdot\frac{S_{NOx}}{K_{NOx}+S_{NOx}}\right)\cdot X_{HB}$ |
| ***Hydrolysis*** | | |
| Aerobic | 15 | $k_{H}\cdot\frac{X_{S}/X_{BH}}{K_{X}+X_{S}/X_{BH}}\cdot\frac{S_{O2}}{K_{O2}^{HB}+S_{O2}}\cdot X_{HB}$ |
| Anoxic | 16 | $k_{H}\cdot\eta_{ANOX}\cdot\frac{X_{S}/X_{BH}}{K_{X}+X_{S}/X_{BH}}\cdot\frac{K_{O2}^{HB}}{K_{O2}^{HB}+S_{O2}}\cdot\frac{S_{NO3}}{K_{NO3}^{HB}+S_{NO3}}\cdot X_{HB}$ |
| Anaerobic | 17 | $k_{H}\cdot\eta_{AN}\cdot\frac{X_{S}/X_{BH}}{K_{X}+X_{S}/X_{BH}}\cdot\frac{K_{O2}^{HB}}{K_{O2}^{HB}+S_{O2}}\cdot\frac{K_{NO3}^{HB}}{K_{NO3}^{HB}+S_{NO3}}\cdot X_{HB}$ |
| *Aeration* | 18 | $K_{La_O2}\cdot(S_{O2}^{*}-S_{O2})$ |
| *Stripping_N$_2$O* | 19 | $K_{La_N2O}\cdot(S_{N2O}-S_{N2O}^{*})$ |
| *Stripping_NO* | 20 | $K_{La_NO}\cdot(S_{NO}-S_{NO}^{*})$ |

| Parameter | Definition | Value | Estimated | CV (%) | Units | Ref. |
|---|---|---|---|---|---|---|
| **AOB** | | | | | | |
| $K_{AOB.NH2OH}$ | S_NH2OH affinity for AOB | 0.3 | | | mgN/L | (1) |
| $K_{AOB.NH2OH.ND}$ | S_NH2OH affinity for AOB during NO reduction | 0.3 | 0.25 | 1.8 | mgN/L | (1) |
| $K_{AOB.NH3}$ | S_NH3 affinity for AOB | 0.0075 | 0.1216 | 3.9 | mgN/L | |
| $K_{AOB.NO.ND}$ | S_NO affinity for AOB | 0.004 | | | mgN/L | (3) |
| $K_{AOB.HNO2}$ | S_HNO2 affinity for AOB | 0.0001 | 0.000672 | 4.4 | mgN/L | (4) |
| $K_{AOB.O2.AMO}$ | S_O2 AMO-mediated affinity constant | 0.4 | 0.225 | 7 | mgO2/L | (5) |
| $K_{AOB.O2.HAO}$ | S_O2 HAO-mediated affinity constant | 0.4 | | | mgO2/L | (5) |
| $K_{AOB.O2.i}$ | S_O2 inhibition constant for AOB | 0.1 | | | mgO2/L | (7) |
| $K_{AOB.i.NH3}$ | S_NH3 inhibition constant for AOB | 10 | | | mgN/L | (6) |
| $K_{AOB.i.HNO2}$ | S_HNO2 inhibition constant for AOB | 0.75 | | | mgN/L | (6) |
| $\varepsilon_{AOB}$ | Reduction factor for HAO-mediated maximum reaction rate | 0.001 | 0.000483 | 1.1 | ( - ) | New |
| $\eta_{NIR}$ | Anoxic reduction factor for NO2 reduction | 0.15 | | | ( - ) | (7) |
| $\eta_{NOR}$ | Reduction factor for NO reduction | 0.15 | 0.157 | 3.2 | ( - ) | (7) |
| $\mu_{AOB.AMO}$ | Maximum AMO-mediated reaction rate | 0.78 | 0.49 | 2 | 1/d | (2) |
| $\mu_{AOB.HAO}$ | Maximum HAO-mediated reaction rate | 0.78 | | | 1/d | (2) |
| $b_{AOB}$ | Endogenous decay rate for AOB | 0.096 | | | 1/d | (2) |
| $Y_{AOB}$ | Yield coefficient for AOB | 0.18 | | | mgCOD/mgN | (2) |
| **NOB** | | | | | | |
| $K_{NOB.HNO2}$ | S_HNO2 affinity for NOB | 0.0001 | | | mgN/L | (2) |
| $K_{NOB.O2}$ | S_O2 affinity constant for NOB | 1.2 | | | mgO2/L | (2) |
| $K_{NOB.i.NH3}$ | S_NH3 inhibition constant for NOB | 0.5 | | | mgN/L | (6) |
| $K_{NOB.i.HNO2}$ | S_HNO2 inhibition constant for NOB | 0.1 | | | mgN/L | (6) |
| $\mu_{NOB}$ | Maximum NOB growth | 0.78 | 0.66 | 1 | 1/d | (2) |
| $b_{NOB}$ | Endogenous decay rate for NOB | 0.096 | | | 1/d | (2) |
| $Y_{NOB}$ | Yield coefficient for NOB | 0.06 | | | mgCOD/mgN | (2) |
| **Others** | | | | | | |
| $f_{XI}$ | Fraction of inerts in biomass | 0.08 | | | ( - ) | (2) |
| $i_{NXB}$ | Nitrogen content of biomass | 0.086 | | | mgN/mgCOD | (2) |
| $i_{NXI}$ | Nitrogen content of inerts | 0.02 | | | mgN/mgCOD | (2) |
| $i_{NXS}$ | Nitrogen content of particulate | 0.06 | | | mgN/mgCOD | (2) |
| $\eta_b$ | Anoxic reduction factor of endogenous decay | 0.33 | | | ( - ) | (5) |
| $K_{O2.b}$ | S_O2 affinity constant of endogenous decay | 0.2 | | | mgO2/L | (5) |
| $K_{NOx}$ | S_NO2+S_NO3 affinity constant of endogenous decay | 0.2 | | | mgN/L | (5) |
| $k_H$ | Hydrolysis rate | 2.21 | 2.01 | 0.9 | 1/d | (2) |
| $K_X$ | Affinity constant for hydrolysis | 0.15 | | | mgCOD/mgC( | (2) |
| $\eta_{anox}$ | Anoxic hydrolysis factor | 0.4 | | | ( - ) | (5) |
| $\eta_{anaer}$ | Anaerobic hydrolysis factor | 0.4 | | | ( - ) | (5) |

| Parameter | Definition | Value | Estimated | CV (% | Units | Ref. |
|---|---|---|---|---|---|---|
| **HB** | | | | | | |
| $K_{HB.NH4}$ | S_NH4 affinity constant for HB | 0.01 | | | mgN/L | (2) |
| $K_{HB.NO3}$ | S_NO3 affinity constant for HB | 0.2 | | | mgN/L | (2) |
| $K_{HB.NO2}$ | S_NO2 affinity constant for HB | 0.2 | | | mgN/L | (2) |
| $K_{HB.NO}$ | S_NO affinity constant for HB | 0.05 | | | mgN/L | (2) |
| $K_{HB.N2O}$ | S_N2O affinity constant for HB | 0.05 | | | mgN/L | (2) |
| $K_{HB.S}$ | S_S affinity constant for HB | 20 | | | mgCOD/L | (2) |
| $K_{HB.S.NAR}$ | S_S affinity constant for S_NO3 reduction | 20 | | | mgCOD/L | (2) |
| $K_{HB.S.NIR}$ | S_S affinity constant for S_NO2 reduction | 20 | | | mgCOD/L | (2) |
| $K_{HB.S.NOR}$ | S_S affinity constant for S_NO reduction | 20 | | | mgCOD/L | (2) |
| $K_{HB.S.NOS}$ | S_S affinity constant for S_N2O reduction | 40 | | | mgCOD/L | (2) |
| $K_{HB.O2}$ | S_O2 affinity constant for HB | 0.1 | | | mgO2/L | (2) |
| $K_{HB.O2.i.NAR}$ | S_O2 inhibition constant for S_NO3 reduction | 0.1 | | | mgO2/L | (2) |
| $K_{HB.O2.i.NIR}$ | S_O2 inhibition constant for S_NO2 reduction | 0.1 | | | mgO2/L | (2) |
| $K_{HB.O2.i.NOR}$ | S_O2 inhibition constant for S_NO reduction | 0.1 | | | mgO2/L | (2) |
| $K_{HB.O2.i.NOS}$ | S_O2 inhibition constant for S_N2O reduction | 0.1 | | | mgO2/L | (2) |
| $K_{HB.NO.i.NIR}$ | S_NO inhibition constant for S_NO2 reduction | 0.5 | | | mgN/L | (2) |
| $K_{HB.NO.i.NOR}$ | S_NO inhibition constant for S_NO reduction | 0.3 | | | mgN/L | (2) |
| $K_{HB.NO.i.NOS}$ | S_NO inhibition constant for S_N2O reduction | 0.075 | | | mgN/L | (2) |
| $\mu_{HB}$ | Maximum HB growth rate | 6.24 | | | 1/d | (2) |
| $\mu_{HB.NAR}$ | Maximum NO3-reduction reaction rate | 1.754 | | | 1/d | (2) |
| $\mu_{HB.NIR}$ | Maximum NO2-reduction reaction rate | 1 | | | 1/d | (2) |
| $\mu_{HB.NOR}$ | Maximum NO-reduction reaction rate | 2.18 | | | 1/d | (2) |
| $\mu_{HB.NOS}$ | Maximum N2O-reduction reaction rate | 2.18 | | | 1/d | (2) |
| $\eta_{HD}$ | Reduction factor for HB denitrification | 0.2 | 0.05 | | ( - ) | ( - ) |
| $b_{HB}$ | Endogenous decay rate for HB | 0.41 | | | 1/d | (2) |
| $Y_{HB}$ | Yield coefficient for HB | 0.6 | | | mgCOD/mgCOD | (2) |

(1) Assumed, (2) Hiatt-Grady 2008, (3) Spérandio 2016, (4) Domingo-Felez 2017, (5) Lisha Guo 2014, (6) Park 2010, (7) Ni 2011,